\documentclass[%
superscriptaddress, %
twocolumn, %
showpacs,
 amsmath,amssymb,
 aps,
prb,
floatfix, longbibliography ]{revtex4-1}
\usepackage{graphicx}
\usepackage{dcolumn}
\usepackage{bm}
\usepackage{soul,color}

\usepackage[colorlinks,linkcolor=blue,anchorcolor=blue,citecolor=blue,urlcolor=blue]{hyperref}
\usepackage{tabularx}

\newcommand{\avg}[1]{\left\langle#1\right\rangle}
\newcommand{\mb}[1]{\mathbf{#1}}
\newcommand{\mr}[1]{\mathrm{#1}}

\begin{document}

\title{Pressure induced gap modulation and topological transitions in twisted bilayer and double bilayer graphene}

\author{Xianqing Lin}
\email[E-mail: ]{xqlin@zjut.edu.cn}
\affiliation{College of Science,
             Zhejiang University of Technology,
             Hangzhou 310023, People's Republic of China}

\author{Haotian Zhu}
\affiliation{College of Science,
             Zhejiang University of Technology,
             Hangzhou 310023, People's Republic of China}

\author{Jun Ni}
\affiliation{State Key Laboratory of Low-Dimensional Quantum Physics and Frontier Science Center for Quantum Information,
             Department of Physics, Tsinghua University, Beijing 100084,
             People's Republic of China}

\date{\today}

\begin{abstract}
We study the electronic and topological properties of fully relaxed twisted bilayer (TBG) and double bilayer (TDBG) graphene
under perpendicular pressure. An approach has been proposed to obtain the equilibrium in-plane structural deformation and out-of-plane
corrugation in  moir\'{e} superlattices under pressure.
We find that the in-plane relaxation becomes much stronger under higher pressure, while the corrugation height in
each layer is maintained.
The comparison between band structures of relaxed and rigid structures demonstrates that
not only the gaps on the electron and hole sides ($\Delta_e$ and $\Delta_h$) are significantly underestimated without relaxation
but also the detailed dispersions of the middle bands of rigid structures are rather different from those of relaxed systems.
$\Delta_e$ and $\Delta_h$ in TBG reach maximum values around critical pressures
with narrowest middle bands.
Topological transitions occur in TDBG under pressure with the middle valence and conduction bands in one valley touching
and their Chern numbers transferred to each other.
The pressure can also tune the gap at the neutrality point of TDBG, which becomes closed for a pressure range and
reopened under higher pressure.
The behavior of electronic structure of supertlattices under pressure is sensitive to the twist angle $\theta$ with
the critical pressures generally increase with $\theta$.
\end{abstract}

\pacs{%
}



\maketitle


\section{Introduction}

Realization of magic-angle twisted bilayer graphene (TBG) has recently intrigued great interest in exploring their peculiar
electronic structure associated with the nearly flat bands around the Fermi level ($E_F$)
\cite{Cao2018,cao2018unconventional,EmergentSharpe605,lu2019superconductors}.
In TBG with twist angles ($\theta$) around the first magic $\theta$ of about 1.1$^\circ$\cite{Bistritzer12233,OriginPhysRevLett.122.106405},
experiments observed correlated-insulator and nonconventional-superconductivity phases\cite{Cao2018,cao2018unconventional,EmergentSharpe605,lu2019superconductors}, which were
found to be extremely sensitive to $\theta$.
With $\theta$ away from the magic angle, the middle bands around $E_F$ become much wider
\cite{Bistritzer12233,OriginPhysRevLett.122.106405,LopesdosSantos2007,Morell2010,Moon2012,Trambly2012,LopesdosSantos2012,Fang2016}, while the perpendicular pressure
has been demonstrated experimentally to be able to flatten these bands again\cite{TuningYankowitz1059}.
Pressure thus provides an efficient way of tuning TBG into the magic regime besides the precise control of $\theta$.
The narrowing of the middle bands under pressure has also been confirmed by theoretical calculations\cite{PressurePhysRevB.98.085144,Pressurechittari2018pressure}.
However, these studies only considered rigid TBG or some relaxation effect in TBG with one empirical parameter
and mainly focused on the widths of middle bands\cite{PressurePhysRevB.98.085144,Pressurechittari2018pressure}, while
the rigid superlattices undergo spontaneous in-plane relaxation and out-of-plane corrugation due to the energy gain from the larger
domains of energetically favorable stacking configurations
\cite{McEuenBLG13,Uchida2014,Wijk2015,Dai2016,Jain2017,Nam2017,Gargiulo2018,carr2018relaxation,ShearPhysRevB.98.195432,Atomicyoo2019atomic,CrucialPhysRevB.99.195419,ContinuumPhysRevB.99.205134}.
The detailed electronic structure of TBG around $E_F$ can be greatly affected by the structural deformation
\cite{Nam2017,Gargiulo2018,ShearPhysRevB.98.195432,CrucialPhysRevB.99.195419,ContinuumPhysRevB.99.205134}, especially the
gaps on the electron and hole sides and the explicit dispersions of the nearly flat bands.
Therefore, it is important to obtain the energetically stable structures of TBG to study the evolution of their electronic structure
under pressure.
First, the variation of realistic structural parameters with pressure can be provided by full relaxation of compressed TBG.
Second, the relaxation effect can be took into account in calculations of the electronic structure of TBG under pressure so that
the critical pressures into the magic regime and the pressure induced modulations of electronic properties corresponding to
experimental systems can be identified.

Besides TBG, twisted double bilayer graphene (TDBG) with relative rotation between the top and bottom graphene bilayers
has been realized recently\cite{Observation2019,Spin-polarizedliu2019,Electriccao2019}. In addition to the nearly flat middle bands in TDBG with a small $\theta$,
a gap at $E_F$ is opened,
and among various stacking arrangements between the bilayers AB-BA TDBG can become valley Hall insulators
\cite{BandPhysRevB.99.235406,FlatPhysRevB.99.235417,IntrinsicPhysRevB.100.201402,QuantumPhysRevX.9.031021,Theorylee2019theory}.
Then the perpendicular pressure may be employed to tune the electronic\cite{FlatPhysRevB.99.235417} and topological properties of TDBG,
and the structural relaxation\cite{IntrinsicPhysRevB.100.201402} can be important to predict these properties under pressure.
It is also noted that the pressure effect on topological properties of TDBG remains to be revealed.

Here we propose an approach to fully relax TBG and AB-BA TDBG under perpendicular pressure.
We find that the in-plane relaxation becomes stronger under higher pressure.
The gaps on the electron and hole sides, the dispersions of middles bands of TBG and the neutrality-point gap of TDBG can
be effectively modulated by pressure. Pressure induced topological transitions can be observed in TDBG.

The outline of this paper is as follows: In Sec. II we study the full
relaxation of moir\'{e} superlattices in TBG and TDBG under perpendicular pressure.
The band structures of systems with and without relaxation are compared in Sec. III.
For the fully relaxed structures, we show the gap modulation and evolution of bands in TBG under pressure in Sec. IV
and the topological and electronic transitions in TDBG induced by pressure in Sec. V.
Section VI presents the summary and conclusions.

\begin{figure*}[t]
\includegraphics[width=1.9\columnwidth]{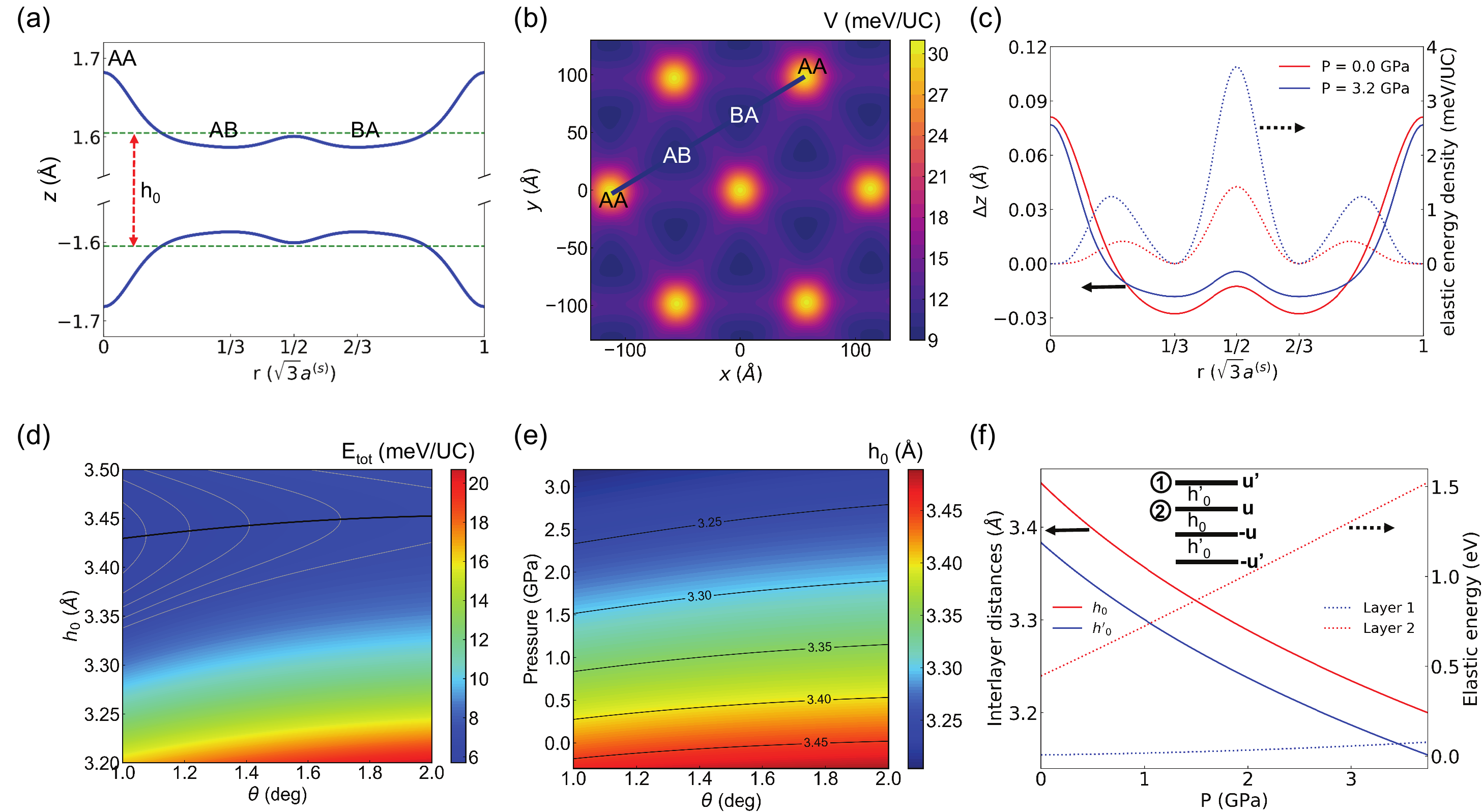}
\caption{(Color online) The full relaxation of TBG and TDBG under perpendicular pressure.
(a) The out-of-plane atomic corrugation of TBG with $\theta = 1.248^{\circ}$ under pressure of 3.2 GPa along the line shown in
(b). The $z$ direction is perpendicular to the layers. $h_0$ is the average interlayer distance and the high-symmetry local stackings are labeled.
(b) The spatial distribution of the local interlayer
interaction energy $V$ as a function of the position in the superlattice
of TBG under pressure of 3.2 GPa. $V$ is in units of meV per graphene unit cell (UC).
(c) The corrugation (solid lines)  and the elastic energy density (dashed lines) in the top layer
of TBG under zero and a high pressure.
(d) The variation of $E_{tot}$ of a TBG supercell with $h_0$ for $\theta$ from 1$^{\circ}$ to 2$^{\circ}$.
$E_{tot}$ is in units of meV per graphene unit cell (UC).
The black line represents the equilibrium $h_0$ of free TBG under zero pressure.
(e) The $h_0$ as a function of the calculated pressure and $\theta$.
(f) The average interlayer distances (solid lines) and the elastic energy (dashed lines) of one layer in the supercell of
TDBG with $\theta = 1.538^{\circ}$ under increasing pressure.
The average interlayer distances between the two middle layers ($h_{0}$) and between the two top layers ($h_{0}'$) and
the displacement fields in the two top layers ($\mb{u}(\mb{r})$ and $\mb{u}'(\mb{r})$) are labeled in the inset.
\label{fig1}}
\end{figure*}

\section{Relaxation of moir\'{e} superlattices under perpendicular pressure}

We consider moir\'{e} superlattices in TBG and AB-BA TDBG, where
the top layer in TBG has been rotated by $\theta$ counterclockwise with the bottom layer fixed
and the top BA-stacked bilayer in TDBG is twisted with respect to
the bottom AB-stacked bilayer. The mathematical formulation of the moir\'{e} superlattice and its reciprocal lattice
are detailed in the Appendix.

The rigid moir\'{e} superlattices undergo spontaneous in-plane relaxation due to the energy gain
from the larger domains of energetically favorable stacking configurations.
Each layer is also corrugated to reach the optimal interlayer distances ($h$)
of the varying stacking configurations across the superlattice.
Here we propose a method to fully relax TBG under perpendicular pressure.
The TBG with a fixed average interlayer distance ($h_0$ shown in Fig. 1(a)) is first relaxed to calculate
the variation of its total energy ($E_{tot}$) with $h_0$. Then
the pressure is obtained by differentiation of $E_{tot}$ with respect to $h_0$.
TDBG under pressure can be relaxed in a similar way.

Among all configurations of the shifted graphene bilayer, the AB and BA stacked structures have the lowest energy, and the
AA stacking is the most unfavorable. In view of the rotation and inversion symmetry relations of
shifted bilayers, the calculated energy of a bilayer with the same $h$ as a function of the shift vector $\bm{\delta}$ can be expressed
as $V(\bm{\delta}, h) = \tilde{V}_0(h) + \tilde{V}(h) \sum_{j=1}^{3} \cos(\mb{G}_{j} \cdot \bm{\delta})$, where the sum is
limited to three shortest vectors, $\mb{G_1} = \mb{b_1}$,
$\mb{G_2} = -\mb{b_1} + \mb{b_2}$, and $\mb{G_3} = -\mb{b_2}$, $\mb{b}_i$ ($i$=1,2) are basis vectors of the reciprocal
lattice of graphene, and components with longer $\mb{G}_{j}$ are negligible\cite{Zhou2015,ShearPhysRevB.98.195432}.
The dependence of $\tilde{V}_0$ and $\tilde{V}$ on $h$ is detailed in the Appendix.
We note that $\tilde{V}(h)$ is positive for $h$ smaller
than 4.22 {\AA} and increases exponentially with decreasing $h$, which indicates that
the high perpendicular pressure applied to TBG with small average $h$ tends to enhance the energy differences among configurations with different stackings.
It can thus be anticipated that
superlattices under higher pressure may have much stronger in-plane structural deformation and also
reduced out-of-plane corrugation, which may indicate stronger influence of relaxation on the
electronic structure.

We have adopted the continuum elastic theory to express $E_{tot}$ of a moir\'{e} supercell.
$E_{tot}$ as the sum of the elastic energy ($E_{el}$) in each layer and
interlayer interaction energy ($E_{int}$) is a functional of the displacement fields
$\mb{u^{(j)}}(\mb{r})$ with $j = 1$ for the bottom layer and $j = 2$ for the top layer and the interlayer distance field $h(\mb{r})$.
The elastic energy in a layer with the displacement field $\mb{u}(\mb{r})$ is given by\cite{Andres2012}
\begin{eqnarray}
&E_{el}&[\mb{u}(\mb{r})] = \int d{\bf{r}} \biggl\{ \frac{\lambda+\mu}{2}
            \left(\frac{\partial u_x}{\partial x} \!+\!
            \frac{\partial u_y}{\partial y}\right)^2
            \nonumber \\
&+&\!\frac{\mu}{2} \left[
         \left(\frac{\partial u_x}{\partial x} -
          \frac{\partial u_y}{\partial y}\right)^2 %
          \!\!\! + \!
         \left(\frac{\partial u_y}{\partial x} +
          \frac{\partial u_x}{\partial y}\right)^2
          \right] \biggr\} ,%
\label{eq1}
\end{eqnarray}
where the integral extends over a moir\'{e} supercell.
We use $\lambda=4.23$~eV/{\AA}$^2$ and $\mu=9.04$~eV/{\AA}$^2$ for the 2D
elastic Lam\'{e} factors~\cite{carr2018relaxation} of graphene.
$E_{int}$ is given by the integral of the local interlayer interaction energy
\begin{equation}
E_{int} = \int V[\bm{\delta}(\mb{r}), h(\mb{r})]d{\mb{r}},
\end{equation}
where $\bm{\delta}(\mb{r}) = \mb{r} - T_{-\theta}\mb{r} + \mb{u}^{(2)}(\mb{r}) - \mb{u}^{(1)}(\mb{r})$ for a relaxed superlattice with
$T_{-\theta}$ denoting the clockwise rotation by $\theta$.
The $\bf{\tilde{u}}^{(n)}(\mb{r})$ and $h(\mb{r})$ have been expanded in Fourier series as
\begin{equation}
\bf{u}^{(n)}({\bf{r}}) = %
\sum_{\bf{G}^{(s)}} \bf{\tilde{u}}^{(n)} ({\bf{G}}^{(s)})
e^{i{\bf{G}}^{(s)}{\cdot}{\bf{r}}}
\end{equation}
and
\begin{equation}
h(\mb{r}) = h_0 + \Delta h(\mb{r}) =
h_0 + \sum_{\bf{G}^{(s)}} \Delta \tilde{h}({\bf{G}}^{(s)})
e^{i{\bf{G}}^{(s)}{\cdot}{\bf{r}}},
\end{equation}
where the summation is over nonzero reciprocal lattice vectors $\bf{G}^{(s)}$ of the supercell and only
$|\mb{G}^{(s)}| \leq 4 |\mb{b}_1^{(s)}|$ are used as $\bf{\tilde{u}}^{(n)}(\mb{r})$ and $h(\mb{r})$
vary smoothly across the supercell.
The average $h(\mb{r})$ over the supercell is just $h_0$.
In the following, we have minimized the total energy functional for a fixed $h_0$
with respect to $\bf{u}^{(n)}({\bf{r}})$ and $\Delta h(\mb{r})$.
With the obtained $E_{tot}$ as a function of $h_0$,
the perpendicular pressure applied to TBG can be evaluated as
\begin{equation}
P = -\frac{1}{\Omega^{(s)}}\frac{d E_{tot}}{dh_0},
\end{equation}
where $\Omega^{(s)}$ is the supercell area
and $\Omega^{(s)} = \sqrt{3}a^2/[8\sin^2(\theta/2)]$ for the considered
moir\'{e} superlattices.
We have extended the method proposed by Nam and
Koshino\cite{Nam2017} to relax TBG with a fixed $h_0$ as detailed in the Appendix.

\begin{figure}[b]
\includegraphics[width=0.9\columnwidth]{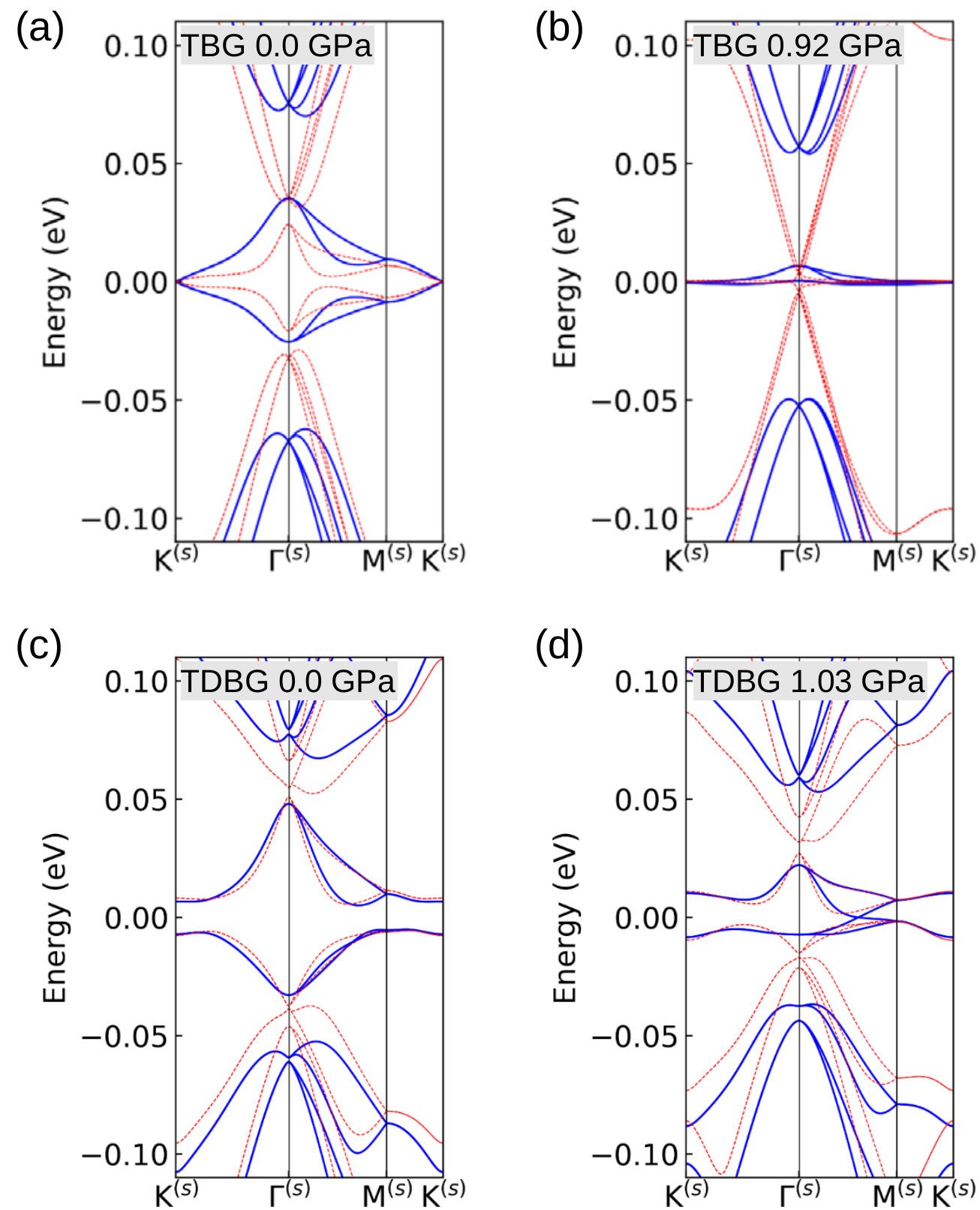}
\caption{(Color online). The band structures of relaxed (solid lines) TBG (a and b) and TDBG (c and d) under zero and a moderate pressure as well as the bands of
rigid superlattices (dashed lines) with interlayer distances the same as the average ones of the relaxed superlattices.
\label{fig2}}
\end{figure}

We find that the layers in TBG under high pressure are still corrugated and
the corrugation height in each layer is just slightly decreased compared with the free structure, as
shown in Figs. 1(a) and 1(c). In contrast, the in-plane relaxation is greatly enhanced by the high pressure
with the AB-like and BA-like parts becoming larger, as seen in Figs. 1(b) and 1(c).
For $\theta = 1.248^{\circ}$, the highest elastic energy density increases by 156\% with pressure
from 0.0 to 3.2 GPa. This is due to the decrease in $h_0$ from 3.44 to 3.21 {\AA},
as the potential energy of the AB-stacking part is much lower than that of the AA-stacking part for the smaller
$h_0$.

Figure 1(d) shows the variation of $E_{tot}$ of TBG with $h_0$ for $\theta$ from 1$^{\circ}$ to 2$^{\circ}$,
from which the pressure $P$ is calculated by Eq. (5). Then
the $h_0$ of systems with given $\theta$ and $P$ can be solved, as seen in Fig. 1(e).
Under the same $P$, $h_0$ increases slowly with $\theta$.
With the same $h_0$, $P$ also increases with $\theta$.

\begin{figure*}[t]
\includegraphics[width=1.9\columnwidth]{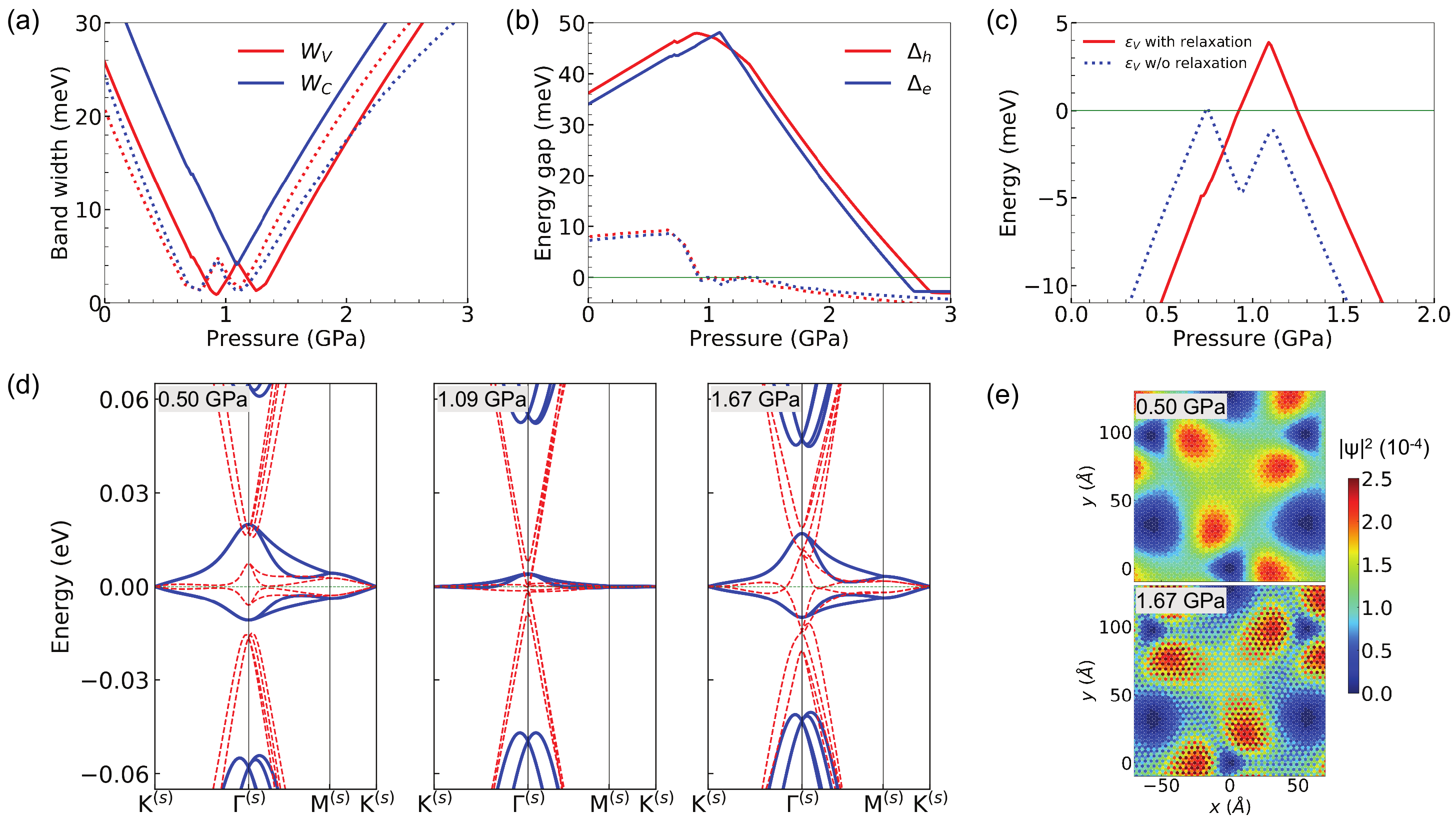}
\caption{(Color online) The evolution of bands with pressure for TBG with $\theta = 1.248^{\circ}$.
(a-c) The widths (a) of middle valence ($W_V$) and conduction bands ($W_C$), the gaps (b) on the electron ($\Delta_e$) and hole ($\Delta_h$)
sides, and the energy (c) of the valence band state at $\Gamma^{(s)}$ with respect to that at $K^{(s)}$
under increasing pressure. The data of relaxed TBG are represented by solid lines and the dashed lines denote rigid
superlattices with interlayer distances the same as the
average ones of the relaxed structures. (d) The band structures of relaxed TBG (solid lines) under pressures of 0.50, 1.09 and 1.67 GPa and those
of corresponding rigid structures (dashed lines). (e) The spatial distribution
at the sublattice-A sites of the bottom layer of the probability density of the middle valence band state at $\Gamma^{(s)}$ for
relaxed TBG under pressures of 0.50 and 1.67 GPa.
\label{fig3}}
\end{figure*}

For TDBG, similar analysis as TBG shows that the displacement fields in the two bottom layers are
the opposite to those in the two top layers ($\mb{u}(\mb{r})$ and $\mb{u}'(\mb{r})$), and the profile of the interlayer distance between the two bottom
layers is the same as that between the two top layers ($h'(\mb{r})$), as shown in Fig. 1(f).
The profile of the interlayer distance between the two middle layers is denoted by $h(\mb{r})$.
The total energy of a TDBG supercell is given by
$E_{tot} = 2 E_{el}[\mb{u}(\mb{r})] + 2 E_{el}[\mb{u}'(\mb{r})] +
\int (V[\bm{\delta}(\mb{r}), h(\mb{r})] +
           2V[\bm{\delta}'(\mb{r}), h'(\mb{r})]) d{\mb{r}}$,
where $\bm{\delta}(\mb{r}) = \mb{r} - T_{-\theta}\mb{r} + 2\mb{u}(\mb{r})$, $\bm{\delta}'(\mb{r}) = \bm{\delta}_{AB} + \mb{u}'(\mb{r}) - \mb{u}(\mb{r})$,
$h(\mb{r}) = h_{0}+\Delta h(\mb{r})$ and $h'(\mb{r}) = h_{0}'+\Delta h'(\mb{r})$.
Here $h_{0}$ and $h_{0}'$ are the average values of $h(\mb{r})$ and $h'(\mb{r})$, respectively.
The TDBG with given $h_{0}$ and $h_{0}'$ is also relaxed by solving Euler-Lagrange equations using the method similar to that for TBG.
The perpendicular pressure applied to TDBG can be obtained as
\begin{equation}
P =  -\frac{1}{\Omega^{(s)}}\frac{\partial E_{tot}}{\partial h_0} =  -\frac{1}{2\Omega^{(s)}}\frac{\partial E_{tot}}{\partial h_0'}.
\end{equation}
Such relations between $h_{0}$, $h_{0}'$  and $P$ ensure the equilibrium of the system so that the internal
pressure applied to each middle layer is equal to the external pressure exerted by the substrates,
which is similar to the equilibrium condition for studies of graphene and boron nitride heterostructures
under pressure\cite{Dynamic2018May}.
It is noted that the proposed approach to relax moir\'{e} superlattices under pressure can be applied to various twisted bilayers and multilayers
including heterostructures.

The obtained fully relaxed TDBG show that the in-plane structural deformation in the top layer and bottom layer is much
smaller than that in the middle layers as the displacement fields in the middle layers do not cause large change of
the AB stacking in the top two layers and the bottom two layers, as shown in Fig. 1(f).
On the contrary, the corrugation in the top layer and bottom layer is similar to that in the middle layers to maintain
the favorable interlayer distance for AB-stacking.
Under pressure, $h_{0}$ is almost equal to that of TBG with the same $\theta$, while
$h'_{0}$ is close to the interlayer distance of AB-stacked BLG under the same pressure. So $h'_{0}$ is smaller than $h_{0}$
with their difference around 0.05 {\AA}.

With the relaxed structures of TBG and TDBG under pressure, their band structures can be obtained based on the tight-binding Hamiltonian.
We will demonstrate below that relaxation is important to describe the pressure induced gap modulation and
topological transitions in TBG and TDBG.

\section{Comparison between band structures of systems with and without relaxation}

To calculate band structures of relaxed TBG and TDBG under pressure,
we extend the Hamiltonian of $p_z$ orbitals for graphene bilayers proposed in Refs.~[\onlinecite{DT268},\onlinecite{ShearPhysRevB.98.195432}] to
take into account the effect of the pressure dependent in-plane structural deformation and
out-of-plane atomic corrugation.
The hopping between intralayer nearest neighbors with their distance $d$ deviated from that of
the pristine graphene $d_0 = a/\sqrt{3}$ is given by
$V_{pp\pi}(d) = -V_{pp\pi}^0 e^{-(d - d_0)/\lambda_{\pi}}$
with $\lambda_{\pi} = 0.47$ {\AA}.
$V_{pp\pi}^0 = 3.09$ eV reproduces the experimental Fermi
velocity\cite{CastroNeto2009} $v_F \approx 1 \times 10^6$ m/s in the graphene layer.
The expression of the interlayer hopping between
sites with in-plane projection $r$ and out-of-plane projection $h$ is
\begin{equation}
V_{pp\sigma}(r, h) = V_{pp\sigma}^0 e^{-(h - h_0)/\lambda'} e^{-(\sqrt{r^2 + h^2} - h)/\lambda} \frac{h^2}{r^2 + h^2},
\end{equation}
where $V_{pp\sigma}^0$ = 0.453 eV, $h_0$ = 3.38 {\AA}, $\lambda'$ = 0.58 {\AA}, and $\lambda$ = 0.27 {\AA}.
All interlayer hopping terms with $r \leq 5.0$ {\AA} are included in the calculations.
We note that these Hamiltonian parameters can reproduce the observed magic angle of free
TBG.

Since TBG and TDBG with small $\theta$ have large moir\'{e} supercells, their Hamiltonian can be
diagonalized using the planewave-like basis functions.
In this approach, the atomic positions of the rigid
graphene lattice in each layer are used to label the hopping sites in the Hamiltonian.
At a k-point $\mb{k}^{(s)}$ in the supercell BZ, a low-energy basis function
for the sublattice $\alpha$ ($\alpha$ = A, B) in layer $n$ with
momentum close to one Dirac point of the layer is defined as
\begin{eqnarray}
|n\alpha, \mb{k}^{(s)} &+& \mb{k}_0 + \mb{G}^{(s)}\rangle =  \nonumber \\
& &\frac{1}{\sqrt{N}} \sum_{\mb{r}_{n\alpha}}
e^{i (\mb{k}^{(s)} + \mb{k}_0 + \mb{G}^{(s)}) \cdot \mb{r}_{n\alpha}}
|\mb{r}_{n\alpha}\rangle,
\end{eqnarray}
where $\mb{k}_0$ is the center
of one of the supercell BZs containing the Dirac points ($K_\xi$ and $K'_\xi$  with $\xi=\pm1$ the valley index) of the fixed and twisted layers at their
corners and $\mb{r}_{n\alpha}$ is the rigid position of a sublattice-$\alpha$ atom in layer $n$.
$\mb{k}_0$ is thus a reciprocal lattice vector of the supercell and
the used $\mb{k}_0$ can be seen in the schematic reciprocal lattice of a moir\'{e} superlattice in Fig. 7(a) of the Appendix.
${\bf{G}^{(s)}}$ is a reciprocal lattice vector of the
superlattice and is given by
${\bf{G}^{(s)}}=j_1~{\bf{b_1^{(s)}}}+j_2~{\bf{b_2^{(s)}}}$ with
small integers $j_1$ and $j_2$ typically in the range of
$-4 \sim 4$.
The Hamiltonian element between two basis functions is given by
\begin{eqnarray}
\langle n\alpha, \mb{k}_1| H |m&&\beta, \mb{k}_2\rangle =
\frac{1}{N_0}  \sum_{\mb{r}_{n\alpha} \in SC} \sum_{\mb{r}_{m\beta}} \nonumber \\
&& e^{-i \mb{k}_1 \cdot \mb{r}_{n\alpha} + i \mb{k}_2 \cdot \mb{r}_{m\beta}}
\langle \mb{r}_{n\alpha} |H| \mb{r}_{m\beta} \rangle,
\end{eqnarray}
where $\mb{k}_j = \mb{k}^{(s)} + \mb{k}_0+\mb{G}^{(s)}_j$ ($j=1,2$),
the summation over $\mb{r}_{n\alpha}$ is done in a supercell, $N_0$ is the number of graphene unit cells in one layer of the supercell,
and only a small number of large hopping terms $\langle \mb{r}_{n\alpha} |H| \mb{r}_{m\beta} \rangle$ are
required for each $\mb{r}_{n\alpha}$.
Since the Hamiltonian between states from the two different valleys is negligible for large
moir\'{e} superlattices, each obtained band can be characterized by the valley index $\xi$,
and this is an advantage of diagonalization of the Hamiltonian using the planewave-like basis functions.

Figure 2 shows the band structures of relaxed TBG and TDBG under zero and a moderate pressure as well as the bands of
rigid superlattices with interlayer distances the same as the average ones of the relaxed superlattices.
The bands without relaxation could give similar band widths as those with relaxation for both TBG and TDBG. However,
the energy gaps above and below the flat bands are significantly underestimated without relaxation
for both TBG and TDBG, and
the gaps at the neutrality point of relaxed TDBG under pressure are rather different from those
without relaxation. Moreover, around the critical pressure with the minimum band width of the middle valence bands,
the dispersions of the middle bands without relaxation are also different from those with relaxation for both systems.
It is also noted that the middle valence and conduction bands touch each other in relaxed TDBG under a moderate
pressure, which may indicate the occurrence of a topological transition.
Therefore, relaxation must be considered to describe the pressure induced gap modulation and
topological transitions in TBG and TDBG.
In comparison, previous studies only focused on pressure induced
narrowing of bands in TBG as full relaxation was not considered there\cite{PressurePhysRevB.98.085144,Pressurechittari2018pressure}.

\begin{figure}[tb]
\includegraphics[width=0.7\columnwidth]{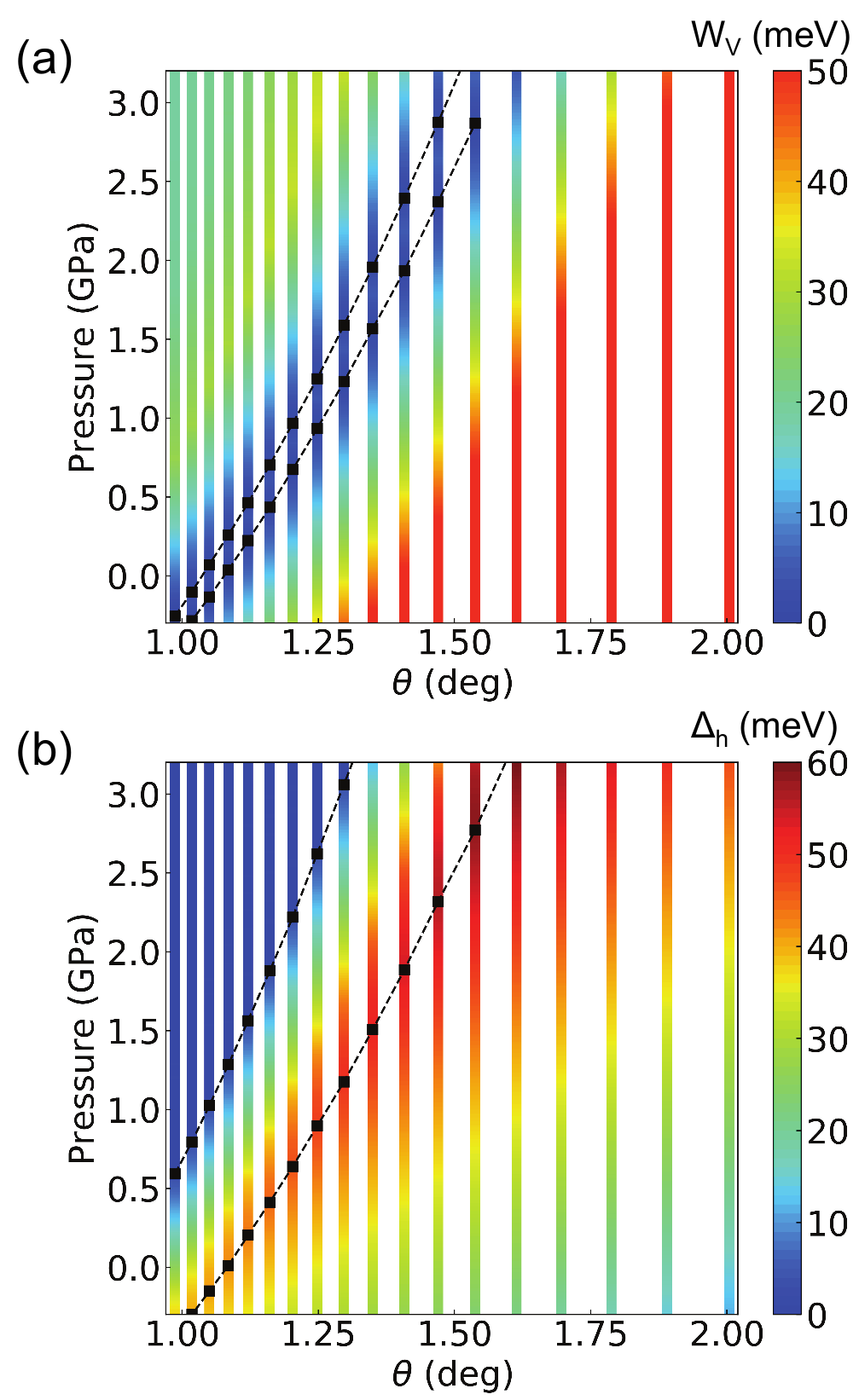}
\caption{(Color online) The $W_V$ (a) and ${\Delta}_h$ (b) as functions of pressure for strictly periodic TBG with
$\theta$ from $1.0^{\circ}$ to $2.0^{\circ}$.
The black squares in (a) show the $P_{c1}$ and $P_{c2}$ for each $\theta$.
The lower and upper squares in (b) show the pressures with maximum and closed ${\Delta}_h$, respectively.
\label{fig4}}
\end{figure}

\begin{figure*}[t]
\includegraphics[width=1.9\columnwidth]{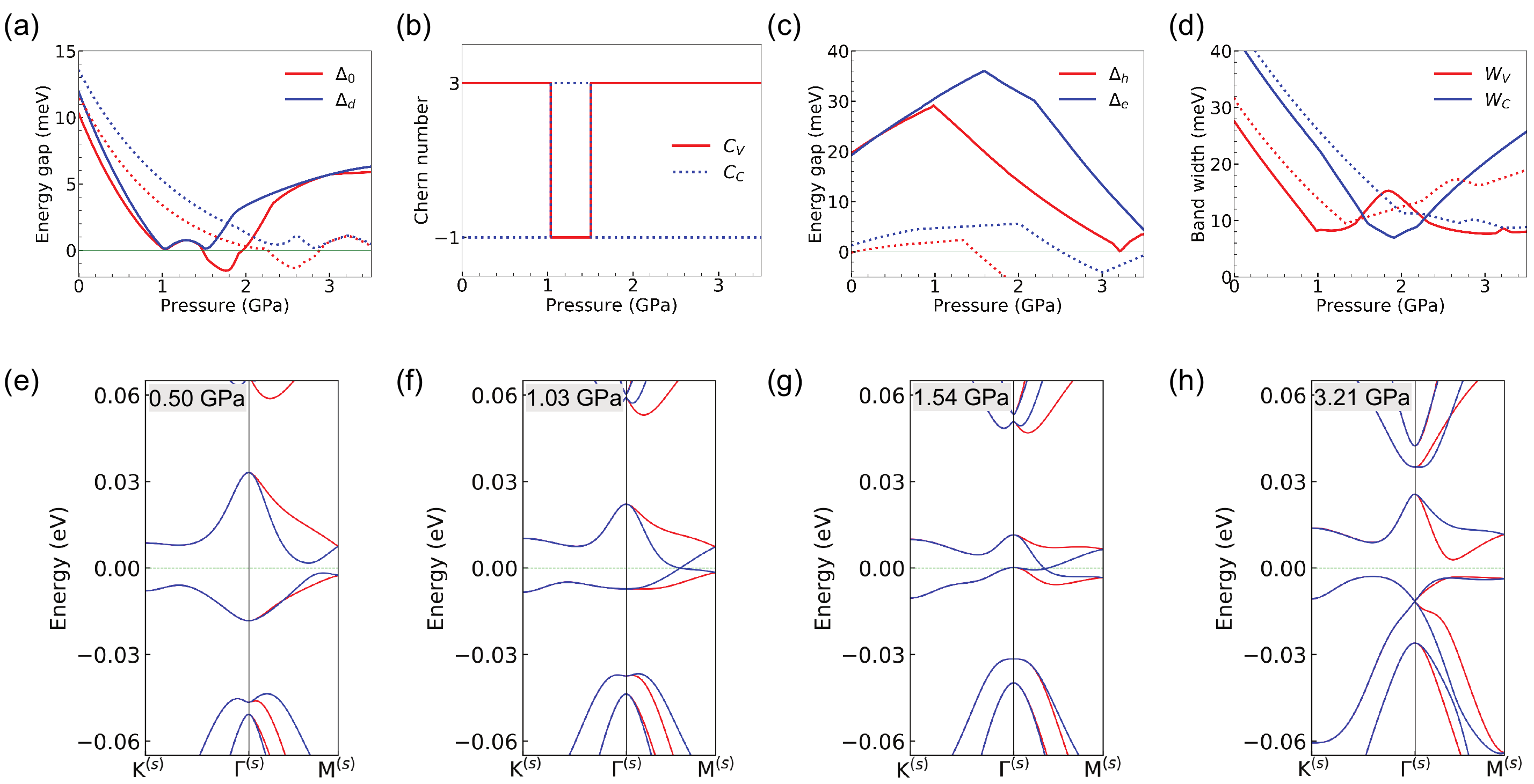}
\caption{(Color online) The variations of electronic and topological properties with pressure for TDBG with $\theta$ = 1.538$^{\circ}$.
(a-d) The gap (a) at the neutrality point ($\Delta_0$) and the minimum direct gap ($\Delta_d$) between the middle valence and conduction bands,
the Chern numbers (b) of the middle valence ($C_V$) and conduction ($C_C$) bands, the gaps (c) on the electron ($\Delta_e$) and hole ($\Delta_h$)
sides, and
the widths (d) of middle valence bands ($W_V$) and conduction bands ($W_C$)
under increasing pressure. The solid and dashed lines represent relaxed and rigid superlattices, respectively in
(a), (c) and (d).
(e-h) The band structures of relaxed TDBG under pressures of 0.50, 1.03, 1.54, and 3.21 GPa.
The blue and red lines represent bands in the $\xi = +$ and $\xi = -$ valleys, respectively.
\label{fig5}}
\end{figure*}

\section{Gap modulation and evolution of bands in TBG by pressure}

At the first magic angle of around 1.1$^{\circ}$, the middle bands at $E_F$ in free TBG flatten.
For $\theta$ larger than the magic angle with wider middle bands, the perpendicular pressure
narrows down these bands.
We find that the pressure can also tune the band gaps on the electron and hole sides, the dispersions of the middle bands as well as the
wavefunctions of states in these bands.
Such electronic structures with full relaxation considered for different $\theta$ and pressure can be based to construct models to
describe the observed correlated and superconducting behavior in TBG under pressure\cite{TuningYankowitz1059}.
We first show the evolution of bands with pressure ($P$) for $\theta = 1.248^{\circ}$.

The width of the middle valence bands ($W_{V}$) reaches local minima at $P_{c1} = 0.92$ GPa and
$P_{c2} = 1.25$ GPa, where $W_V$ is smaller than 2 meV, as shown in Fig. 3(a).
Due to the absence of particle-hole symmetry in relaxed systems, the width of the middle conduction bands ($W_C$) is always larger
than $W_V$ with $W_C$ reaching the minimum value of 4 meV under pressure between $P_{c1}$ and $P_{c2}$.
We note that experiments have shown that the middle bands in TBG with $\theta = 1.27^{\circ}$
become flat under the pressure of 1.33 GPa\cite{TuningYankowitz1059}, which is just slightly larger than the calculated $P_{c2}$.
$W_V$ and $W_C$ of rigid superlattices become minimum at similar $h_0$ as those of relaxed structures.
With the narrowing of middle bands under pressure, the gaps on the electron ($\Delta_e$) and hole ($\Delta_h$)
sides are enhanced by the pressure and reach maximum values under pressure around $P_{c1}$ and $P_{c2}$,
as shown in Fig. 3(b).
When $P$ is larger than the critical pressures, $\Delta_e$ and $\Delta_h$ decrease with $P$ and become
closed under $P\approx 2.8$ GPa. In contrast, the $\Delta_e$ and $\Delta_h$ of rigid superlattices are already
closed with $h_0$ around that of the relaxed structure under $P\approx 1.0$ GPa.

Under pressure between $P_{c1}$ and $P_{c2}$, we find that the nearly flat valence and conduction bands become overlapped as indicated
by the higher energy of the valence band state at $\Gamma^{(s)}$ than the degenerate valence and conduction states
at $K^{(s)}$, as shown in Figs. 3(c) and 3(d).
The varying band structures with $P$ can also be seen
in video 1 of the Supplemental Material (SM)\cite{press-SM}.
With pressure increasing from $P_{c1}$ to $P_{c2}$, the flat valence bands around $\Gamma^{(s)}$ move up while
the flat conduction bands around $\Gamma^{(s)}$ move down, and they switch at $P = 1.09$ GPa.
Such evolution of band dispersions may suggest a transition of the electronic structure.
It is noted that under pressure beyond $P_{c1}$ and $P_{c2}$, the bands around $E_F$ are just those
with linear dispersions around $K^{(s)}$ (see Fig. 3(d)), so the density of state (DOS) at $E_F$ is zero.
However, when the flat valence and conduction bands overlap, the DOS at $E_F$ can be very large.
In addition, the middle bands under a large pressure may be similar to those under a small pressure (see Fig. 3(d)),
while the wavefunctions of states around $\Gamma^{(s)}$ are distinct from each other, especially the positions
with highest probability density,
as shown in Fig. 3(e).

Since the electronic properties of TBGs are sensitive to $\theta$,
the effect of pressure on tuning their band structures also
depends closely on $\theta$.
$P_{c1}$ and $P_{c2}$ increase rapidly with $\theta$, and
only for $\theta$ smaller than about 1.54$^{\circ}$ $W_V$ can reach the minimum
under $P \leq 3.2$ GPa, as shown in Fig. 4(a).
In addition, the pressure range $P_{c2} - P_{c1}$ with overlapped valence and conduction bands
increases with $\theta$.
For $\theta$ larger than about 1.3$^{\circ}$, $\Delta_e$ and $\Delta_h$ remain open
under $P \leq 3.2$ GPa, and the maximum $\Delta_h$ increases with $\theta$ though
$\Delta_h$ of free systems decreases with $\theta$, as shown in Fig. 4(b).
With $\theta$ from 1.08$^{\circ}$ to 1.54$^{\circ}$, the maximum $\Delta_h$ increases from
42 to 59 meV.

\section{Topological and electronic transitions in TDBG under pressure}

\begin{figure}[t]
\includegraphics[width=1.0\columnwidth]{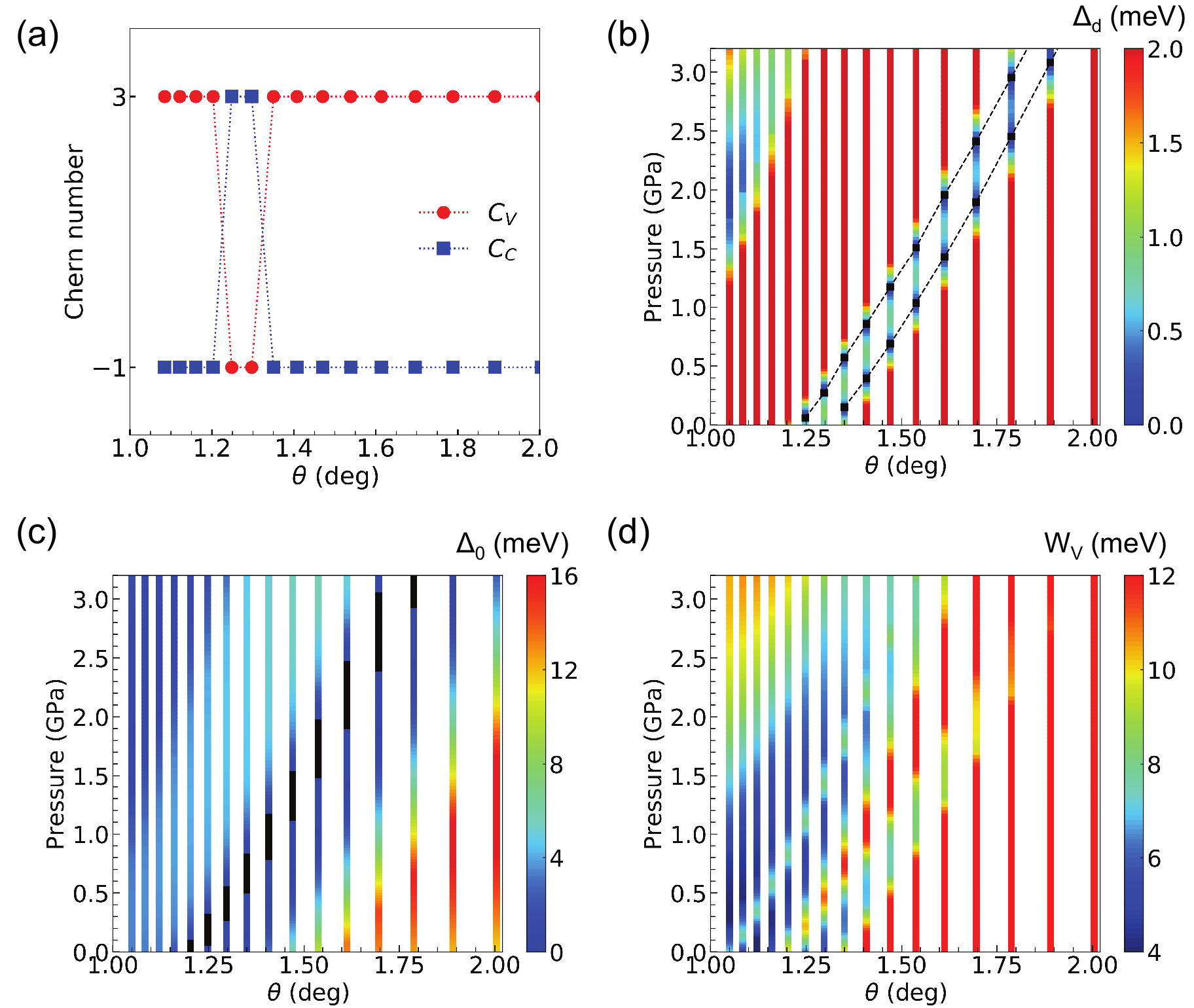}
\caption{(Color online).
(a) The Chern numbers of the middle valence ($C_V$) and conduction ($C_C$) bands for
free TDBGs with varying $\theta$ from 1.0$^{\circ}$ to 2.0$^{\circ}$.
(b-d) The $\Delta_d$ (b), $\Delta_0$ (c) and $W_V$ (d) as functions of pressure for strictly periodic TDBG with
$\theta$ from $1.0^{\circ}$ to $2.0^{\circ}$.
The black squares in (b) show the $P_{c1}$ and $P_{c2}$ for each $\theta$.
The black areas in (c) denote the pressure range with closed $\Delta_0$.
\label{fig6}}
\end{figure}

Free AB-BA TDBGs were found to be valley Hall insulators with a gap ($\Delta_0$) opened at the charge
neutrality point\cite{BandPhysRevB.99.235406,FlatPhysRevB.99.235417}. We observe topological transitions
in TDBG induced by the perpendicular pressure
with the Chern number of the middle valence band transferred to the middle conduction band in a valley.
Electronic transitions with $\Delta_0$ closed and reopened also occur under pressure.

The topological properties of TDBG are characterized by Chern numbers of the bands in each valley, which are
defined as $C_n = 1/2\pi \int \mathcal{F}_{n,\mb{k}^{(s)}} d^2\mb{k}^{(s)}$,
where $\mathcal{F}_{n,\mb{k}^{(s)}}$ is the Berry curvature of the state at $\mb{k}^{(s)}$ in the band with index $n$\cite{RevModPhys.82.1959}.
Due to the absence of inversion symmetry, states in TDBG can have nonzero Berry curvature.
We have calculated $\mathcal{F}_{n,\mb{k}^{(s)}}$ of a relaxed superlattice by\cite{RevModPhys.82.1959}
\begin{eqnarray}
  &&\mathcal{F}_{n,\mb{k}^{(s)}} = -2\ \mr{Im} \sum_{n'\neq n} \nonumber \\
  && \frac{ \avg{n\mb{k}^{(s)}| \frac{\partial H_{\mb{k}^{(s)}}}{\partial k^{(s)}_x} | n'\mb{k}^{(s)} }
  \avg{n'\mb{k}^{(s)}|\frac{\partial H_{\mb{k}^{(s)}}}{\partial k^{(s)}_y} | n\mb{k}^{(s)} } }
                          {(\varepsilon_{n\mb{k}^{(s)}} - \varepsilon_{n'\mb{k}^{(s)}})^2 },
\end{eqnarray}
where the band state $| n\mb{k}^{(s)} \rangle$ and $\partial H_{\mb{k}^{(s)}}/\partial k^{(s)}_\gamma$ ($\gamma = x, y$) are represented in the Bloch basis
$| \nu\mb{k}^{(s)} \rangle = 1/\sqrt{N_s} \sum_{\mb{R}^{(s)}} e^{i\mb{k}^{(s)} \cdot (\mb{R}^{(s)} + \mb{r}_\nu)} | \mb{R}^{(s)} + \mb{r}_\nu \rangle$
with $\mb{r}_\nu$ denoting the rigid position of a site in the supercell and $\mb{R}^{(s)}$ a superlattice vector.
The band state $| n\mb{k}^{(s)} \rangle$ expanded as $\sum_{m\alpha, \mb{G}^{(s)}} \psi_{m\alpha, \mb{G}^{(s)}} |m\alpha, \mb{k}^{(s)} + \mb{k}_0 + \mb{G}^{(s)}\rangle$
can be represented in the basis of $| \nu\mb{k}^{(s)} \rangle$ as
$\sum_{\nu} \left(\sum_{m\alpha, \mb{G}^{(s)}} \psi_{m\alpha, \mb{G}^{(s)}} 1/\sqrt{N_0} e^{i( \mb{k}_0 +  \mb{G}^{(s)}) \cdot \mb{r}_\nu} \right)
| \nu\mb{k}^{(s)} \rangle$.
The matrix representation of $\partial H_{\mb{k}^{(s)}}/\partial k^{(s)}_\gamma$ is
\begin{eqnarray}
\left(\frac{\partial H_{\mb{k}^{(s)}}}{\partial k^{(s)}_\gamma} \right)_{\mu\nu} = \sum_{\mb{R}^{(s)}}& &e^{i\mb{k}^{(s)}
         \cdot (\mb{R}^{(s)} + \mb{r}_\nu - \mb{r}_\mu)} i(\mb{R}^{(s)} + \mb{r}_\nu - \mb{r}_\mu)_\gamma \nonumber \\
         & & \avg{\mb{r}_\mu|H|\mb{R}^{(s)} + \mb{r}_\nu}
\end{eqnarray}
with $\avg{\mb{r}_\mu|H|\mb{R}^{(s)} + \mb{r}_\nu}$ the Hamiltonian between sites at $\mb{r}_\mu$ and $\mb{R}^{(s)} + \mb{r}_\nu$.

We first show the evolution of topological and electronic properties with pressure
for TDBG with $\theta$ = 1.538$^{\circ}$, whose total Chern number of the valence bands is 2 in the
$\xi = +$ valley\cite{BandPhysRevB.99.235406}.

Since the topological transition occurs when the middle valence and conduction bands touch each other,
the minimum direct gap ($\Delta_d$) is calculated to identify the transitions.
The Chern numbers of the middle valence and conduction bands of the free system in the
$\xi = +$ valley are 3 and -1, respectively.
Under the critical pressures $P_{c1} = 1.03$ GPa and $P_{c2} = 1.51$ GPa, the valence and conduction bands touch
and their Chern numbers are transferred to each other, as shown in Figs. 5(a), 5(b), 5(f), and 5(g).
Then under pressure between $P_{c1}$ and $P_{c2}$, the
total Chern number of the valence bands becomes -2. In the $\xi = -$ valley, the Chern number of each band is just
the opposite of that in the $\xi = +$ valley.
For the rigid superlattice, the valence and conduction bands only touch under a very high pressure, as shown in Fig. 5(a).

Besides topological properties, $\Delta_0$ and the gaps on the electron and hole sides ($\Delta_e$
and $\Delta_h$) are also tuned by the pressure.
Under pressure from 0 to $P_{c1}$, $\Delta_0$ decreases to zero. $\Delta_0$ becomes negative with overlapped
valence and conduction bands under pressure slightly larger than $P_{c2}$, the system thus becomes metallic in this pressure range.
Then $\Delta_0$ becomes positive under increasing pressure, as shown in Figs. 5(a) and 5(h).
The varying band structures with pressure can also be seen in
video 2 of the Supplemental Material (SM)\cite{press-SM}.
The trend of $\Delta_e$ and $\Delta_h$ with pressure is similar to that of TBG.
Under pressure of about 3.2 GPa, the middle valence bands touch the lower valence bands, while $\Delta_h$ becomes reopened under
increasing pressure, as shown in Figs. 5(c) and 5(h).

The trend of $W_{V}$ and $W_{C}$ with pressure is shown in Fig. 5(d).
At about $P_{c1}$, $W_{V}$ reaches the minimum value of 10 meV, while $W_{C}$
only becomes minimum under a much higher pressure of 1.91 GPa.
In contrast to the increasing $W_{V}$ beyond the critical pressures in TBG,
the $W_{V}$ of TDBG can become small under high pressure.

Similar topological transitions as those under
increasing pressure also occur in free TDBGs with varying $\theta$ from 2.0$^{\circ}$ to 1.0$^{\circ}$,
as shown in Fig. 6(a).
For $\theta$ from 1.248$^{\circ}$ to 1.890$^{\circ}$, topological transitions can be induced by pressure smaller than 3.2 GPa,
as shown in Fig. 6(b).
In particular, the critical pressures are close to 0 GPa for $\theta = 1.248^{\circ}$.
For $\theta$ smaller than 1.248$^{\circ}$, no topological transitions are observed.
The semiconductor to metal transition under pressure can be observed for $\theta >= 1.203^{\circ}$,
while $\Delta_0$ remains positive and decreases slowly for a smaller $\theta$,
as shown in Fig. 6(c).
Figure 6(d) shows variations of $W_{V}$ with pressure for different $\theta$.
For small $\theta$, the profile of $W_{V}$ can be rather complicated.
We note that the pressure induced minimum $W_{V}$ tends to decrease with $\theta$
from 2.0$^{\circ}$ toward 1.0$^{\circ}$.
For $\theta = 1.050^{\circ}$, the minimum $W_{V}$ is 3.89 meV, while it becomes 7.65 meV for
$\theta = 1.538^{\circ}$.

\section{Summary and Conclusions}

An approach has been proposed to fully relax TBG and TDBG under perpendicular pressure with
both in-plane structural deformation and out-of-plane corrugation considered.
The moir\'{e} superlattices with fixed average interlayer distances $h_0$ are first relaxed and
the pressure is obtained through the variation of the total energy with $h_0$.
We find that the in-plane relaxation is greatly enhanced by the high pressure in TBG and in the middle layers
of TDBG, while the corrugation height in each layer is just slightly decreased by high pressure.
The electronic and topological properties of relaxed superlattices under pressure have been explored
by diagonalizing the tight-binding Hamiltonian using the planewave-like basis functions, so that
each band can be conveniently characterized by the valley index $\xi$.
The comparison between band structures with and without relaxation demonstrates that
relaxation is required to describe the pressure dependent band gaps and detailed band dispersions.
Only band widths without relaxation are similar to those of relaxed systems.

In TBG, the gaps on the electron and hole sides reach maximum values around the critical pressures $P_{c1}$ and $P_{c2}$
with narrowest middle bands and become closed only under a very high pressure.
The nearly flat valence and conduction bands become overlapped between $P_{c1}$ and $P_{c2}$,
and the wavefunctions of band states under higher pressures than $P_{c2}$ exhibit
rather different spatial distributions from those under smaller pressures.
Topological transitions occur in AB-BA TDBGs under pressure with the middle valence and conduction bands in one valley touching
and their Chern numbers transferred to each other.
The pressure can also tune the gap $\Delta_0$ at the neutrality point of TDBG, which becomes closed for a pressure range and
reopened under higher pressure.
The behavior of the electronic structure of supertlattices under pressure is sensitive to the twist angle $\theta$ with
the critical pressures generally increasing with $\theta$.
Our study thus provides a systematic description of electronic and topological properties of fully relaxed TBG and TDBG under pressure,
and the proposed approach to relax moir\'{e} superlattices under pressure can be applied to various twisted bilayers and multilayers under pressure
including heterostructures.

\label{Acknowledgments}
\begin{acknowledgments}
We gratefully acknowledge valuable discussions with D. Tom\'anek, D. Liu,
H. Xiong, and Q. Zhang.
This research was supported by
the National Natural Science Foundation of China (Grant Nos. 11974312 and 11774195),
and the National Key Research and Development Program of China(Grant No. 2016YFB0700102).
\end{acknowledgments}

\section*{Appendix}
\setcounter{equation}{0}
\renewcommand{\theequation}{A\arabic{equation}}

\subsection{Mathematical formulation of the moir\'{e} superlattices}

\begin{figure}[tb]
\includegraphics[width=1.0\columnwidth]{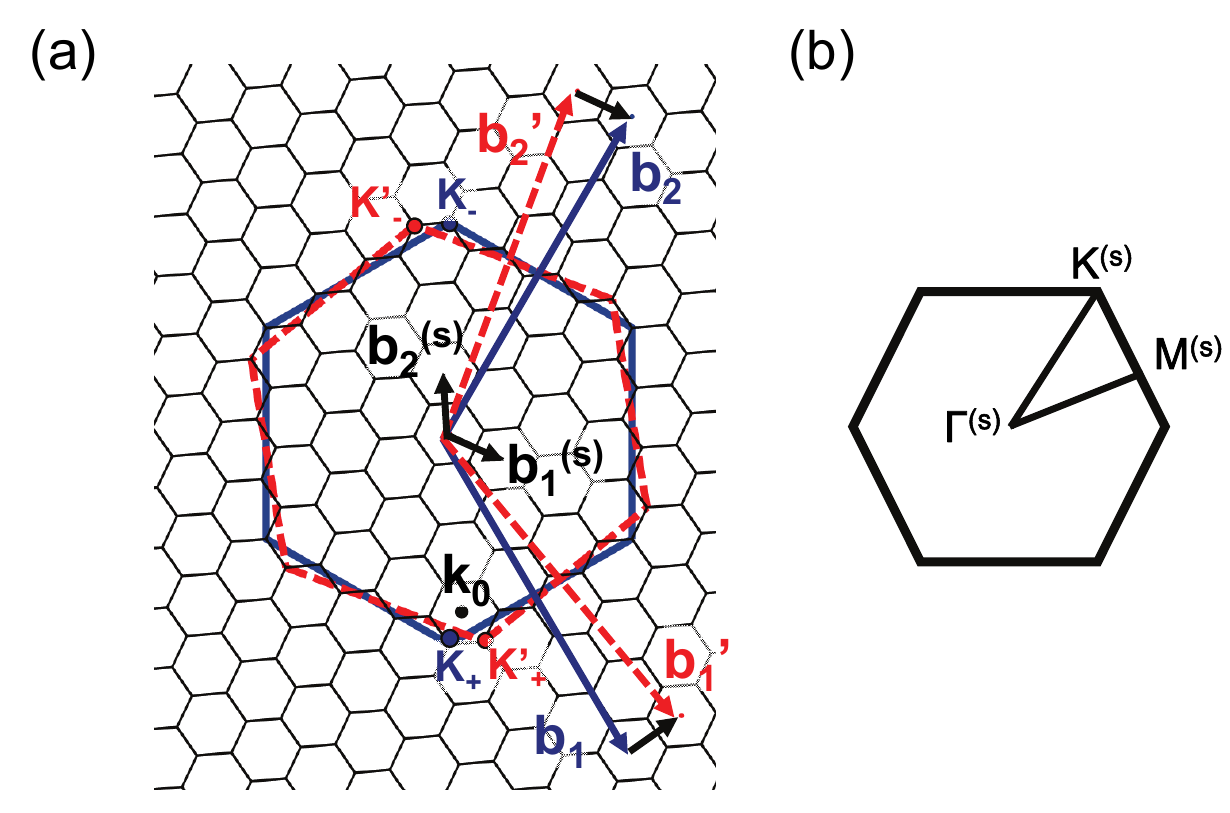}
\caption{(Color online).
(a) The schematic reciprocal lattice of the moir\'{e} superlattice in TBG and TDBG.
Small hexagons are periodic BZs of the superlattice, spanned by
${\bf{b_1^{(s)}}}$ and ${\bf{b_2^{(s)}}}$.
Large hexagons are BZ of the fixed layers (blue), spanned by $\bf{b}_1$
and $\bf{b}_2$, and BZ of the twisted layers (red), spanned by
$\bf{b}'_1$ and $\bf{b}'_2$.
The Dirac points of the fixed ($K_{\xi}$) and twisted ($K'_{\xi}$) layers
are located at corners of the supercell BZs, with $\xi=\pm1$ the valley index.
$\mb{k}_0$ is the center
of one of the supercell BZs containing $K_\xi$ and $K'_\xi$ at their
corners. (b) The high-symmetry k-points in the supercell BZ.
\label{fig7}}
\end{figure}

In the TBG moir\'{e} superlattice, the top layer has been rotated
by $\theta$ counterclockwise with respect to the fixed bottom layer,
and one sublattice-A atom in the bottom layer is placed just below one
sublattice-A atom in the top layer at the origin.
In the considered AB-BA TDBG, the top BA-stacked bilayer is twisted with
the bottom AB-stacked bilayer fixed, and
one sublattice-A atom in each of the two middle layers is placed at the origin.

The unit cell of a fixed layer is spanned by the basis vectors $\mb{a_1} = a(\sqrt{3}/2, -1/2)$ and
$\mb{a_2} = a(\sqrt{3}/2, 1/2)$, where $a = 2.46$ {\AA} is the lattice constant
of graphene.
Then the basis vectors of a twisted layer become
$\mb{a'_j} = T_\theta \mb{a_j}$ ($j$ = 1, 2), where $T_\theta$ denotes the rotation.
The sublattice A atom in a unit cell of each layer is located at
the origin of the cell, and the positions of the sublattice B atoms in the
unit cell of the fixed and twisted layers are $(\mb{a_1} + \mb{a_2})/3$ and $(\mb{a'_1} + \mb{a'_2})/3$, respectively.

We consider strictly periodic moir\'{e} superlattices with basis vectors
$\mb{a_1^{(s)}} = N \mb{a_1} + (N+1) \mb{a_2}$ and $\mb{a_2^{(s)}} = -(N+1) \mb{a_1} + (2N+1) \mb{a_2}$, where $N$ is an integer.
The relation between $\theta$ and $N$ can be expressed as
$\cos \theta = (1+6N+6N^2)/(2+6N+6N^2)$. For $N$ from 16 to 32, $\theta$
takes values from 2.005$^\circ$ to 1.018$^\circ$.
The reciprocal lattice of such a moir\'{e} superlattice is
spanned by the vectors ${\bf{b_1^{(s)}}} =
{\bf{b}_2}-{\bf{b}'_2}$, and ${\bf{b_2^{(s)}}} =
({\bf{b}'_1}+{\bf{b}'_2})-({\bf{b}_1}+{\bf{b}_2})$, where
${\bf{b}_i}$ and ${\bf{b}'_i}$ with $i=1,2$ are reciprocal lattice
vectors of the fixed and twisted layers, respectively, as shown in Fig. 7(a).
The Dirac points of the fixed and twisted layers ($K_{\xi}$ and $K'_{\xi}$ with $\xi=\pm1$ the valley index) are just
located at two adjacent corners of a supercell Brillouin zone (BZ), whose center is denoted by
$\mb{k}_0$. The $\mb{k}_0$ insider the large BZs in Fig. 7(a) is just $-N \bf{b^{(s)}_2}$.
The high-symmetry k-points in the supercell BZ are labeled in Fig. 7(b).

In a large moir\'{e} superlattice, the local lattice structure around a position $\mb{r}$
is similar to that of a shifted graphene bilayer, where
the shift vector ($\bm{\delta}$) is defined as the in-plane displacement
vector from an atom in the bottom layer to the atom in the top layer
which sits just above the bottom-layer atom for the AA stacked bilayer
with $\bm{\delta} = 0$.
Then we find that the local $\bm{\delta}$ between the twist and fixed layers at $\mb{r}$ in the superlattice
can be taken as $\bm{\delta} = (I - T_{-\theta}) \mb{r}$. The $\bm{\delta}$ at $\mb{r} = \mb{a_1^{(s)}}$
and $\mb{a_2^{(s)}}$
are just $\mb{a_2}-\mb{a_1}$ and $-\mb{a_1}$, respectively, which is consistent with the AA stacking at the superlattice vectors of TBG.

\subsection{Dependence of energies of shifted graphene bilayers on the interlayer distance}

The dependence of $\tilde{V}_0$ and $\tilde{V}$ on $h$
can be obtained from the variation of $V$ with $h$ for the
AA- and AB-stacked bilayers ($V_{AA}(h)$ and $V_{AB}(h)$) with $\bm{\delta}_{AA} = 0$ and $\bm{\delta}_{AB} = (\mb{a}_1 + \mb{a}_2)/3$
as $\tilde{V}_0(h) = [V_{AA}(h) + 2V_{AB}(h)]/3$ and $\tilde{V}(h) = 2[V_{AA}(h) - V_{AB}(h)]/9$.
By \emph{ab initio} calculations of energies of bilayers, $V_{AA}(h)$
and $V_{AB}(h)$ are fitted as
$V_{AA}(h) = 0.0322\ e^{-(h - 3.2)/0.316} - (1.443/h)^4 + 0.0205$
and $V_{AB}(h) = 0.0259\ e^{-(h - 3.2)/0.344} - (1.488/h)^4 + 0.0222$
with $V$ and $h$ in units of eV/{\AA}$^2$ and {\AA}, respectively, where
the first terms represent the short-range repulsion and the second terms represent
the long-range van der Waals interaction\cite{Zhou2015}.
From $V_{AA}(h)$ and $V_{AB}(h)$, the estimated corrugation height in free TBG is about 0.12 {\AA},
which is slightly smaller than that of about 0.14 {\AA} from direct \emph{ab initio} relaxation
of magic-angle TBG\cite{CrucialPhysRevB.99.195419}.

We use \emph{ab initio} density functional theory (DFT) as implemented in
the VASP code\cite{{VASP1},{VASP2},{VASP3}} to calculate energies of shifted graphene bilayers with different
interlayer distances.
We use projector augmented-wave
pseudopotentials\cite{{PAW1},{PAW2}} and the SCAN+rVV10 exchange correlation
functional, which provides a proper description
of the van der Waals interaction\cite{Klime2011,Peng2016}.
We use 800 eV
as the kinetic energy cutoff for the plane-wave
basis and the tolerance
for the energy convergence is 10$^{-6}$ eV.
The BZ sampling
is done using a 36 $\times$ 36 $\times$ 1 Monkhorst-Pack grid\cite{SPECIAL1976}, and
vacuums in the z direction are larger than 17 {\AA}.

\subsection{Relaxation of TBG with a fixed average interlayer distance by Euler-Lagrange equations}

We have extended the method proposed by Nam and
Koshino\cite{Nam2017} to relax TBG with a fixed $h_0$.
The $E_{tot} = \sum_{n=1}^2 E_{el}[\mb{u}^{(n)}] + E_{int}$ of a supercell
can be expressed as
$E_{tot} = \int L[\mb{u}^{(1)}, \mb{u}^{(2)}, h_0 + \Delta h] d{\mb{r}}$.
The minimization of $E_{tot}$ as a functional of $\mb{u}^{(n)}$ and $\Delta h$ leads to
a serial of Euler-Lagrange equations
\begin{equation}
\frac{\partial}{\partial x}\left[\frac{\partial L}{\partial (\partial u^{(n)}_\nu/\partial x)}\right] +
\frac{\partial}{\partial y}\left[\frac{\partial L}{\partial (\partial u^{(n)}_\nu/\partial y)}\right] - \frac{\partial L}{\partial u^{(n)}_\nu} = 0
\end{equation}
and
\begin{equation}
\frac{\partial L}{\partial \Delta h} = \frac{\partial V({\bm{\delta}}, h_0+\Delta h)}{\partial \Delta h}  = 0\,,
\end{equation}
where $n = 1, 2$ and $\nu = x, y$. With $\partial L/\partial \mb{u}^{(2)} =
-\partial L/\partial \mb{u}^{(1)} = \partial V/\partial \bm{\delta}$ expanded as
$\sum_{\mb{G^{(s)}}} \mb{\tilde{f}}(\mb{G^{(s)}}) e^{i \mb{G^{(s)}} \cdot \mb{r}}$, substitution of the
Fourier expansion of $\mb{u}^{(n)}(\mb{r})$ into Eq. (4) leads to
\begin{eqnarray}
\left(\begin{matrix}
   (\lambda+2\mu)q^2_x + \mu q^2_y  & (\lambda+\mu)q_x q_y \cr
   (\lambda+\mu)q_x q_y & (\lambda+2\mu)q^2_y + \mu q^2_x \cr
   \end{matrix} \right)
   \left(\begin{matrix}
   u^{(n)}_x(\mb{q}) \cr
   u^{(n)}_y(\mb{q}) \cr
   \end{matrix} \right) = \nonumber \\
   (-1)^{n-1}
   \left(\begin{matrix}
   \tilde{f}_x(\mb{q}) \cr
   \tilde{f}_y(\mb{q}) \cr
   \end{matrix} \right),\ \ \
\label{eqS1}
\end{eqnarray}
where $\mb{q}$ takes each $\mb{G^{(s)}}$.
It can be inferred from this equation that $\mb{u}^{(1)}(\mb{r}) = -\mb{u}^{(2)}(\mb{r})$.
Expanding $\partial V({\bm{\delta}}, h_0+\Delta h)/\partial \Delta h$ as
$\sum_{\bf{G}^{(s)}} \tilde{g}({\bf{G}}^{(s)})
e^{i {\bf{G}}^{(s)}{\cdot}{\bf{r}}}$, Eq. (7) gives rise to
\begin{equation}
\tilde{g}({\bf{G}}^{(s)}) = 0.
\end{equation}
In view of the
real-valued fields and the 120$^\circ$-rotation symmetry of the superlattice,
$\mb{\tilde{u}^{(n)}}(C_{3z}\mb{G^{(s)}}) = C_{3z} \mb{\tilde{u}^{(n)}}(\mb{G^{(s)}})$,
$\mb{\tilde{u}^{(n)}}(-\mb{G^{(s)}}) = \mb{\tilde{u}^{(n)*}}(\mb{G^{(s)}})$,
$\Delta \tilde{h}(C_{3z}\mb{G^{(s)}}) = \Delta \tilde{h}(\mb{G^{(s)}})$,
and $\Delta \tilde{h}(-\mb{G^{(s)}}) = \Delta \tilde{h}^*(\mb{G^{(s)}})$.
So only Fourier components for one sixth of the considered $\mb{G^{(s)}}$
are independent in these equations.
We have solved Eqs. (8) and (9) self consistently
to obtain converged
$\mb{\tilde{u}^{(n)}}(\mb{G^{(s)}})$ and $\Delta \tilde{h}(\mb{G^{(s)}})$ using zero as their initial values.
During the self-consistent iteration, all $\Delta \tilde{h}({\bf{G}}^{(s)})$ remain real so
$\Delta h(\mb{r})$ can be expressed as $\sum_{\bf{G}^{(s)}} \Delta \tilde{h}({\bf{G}}^{(s)})
\cos({\bf{G}}^{(s)}{\cdot}{\bf{r}})$, which indicates that the AA-stacked part at $\mb{r}=0$ has
the largest $h$. Since $\partial V/\partial \bm{\delta}$
remains an odd function of $\mb{r}$, all $\mb{\tilde{u}^{(n)}}(\mb{G^{(s)}})$ are
purely imaginary and $\mb{\tilde{u}^{(n)}}(\mb{r})$ can be expressed as
$-\sum_{\bf{G}^{(s)}} \mr{Im}[\mb{\tilde{u}^{(n)}}({\bf{G}}^{(s)})]
\sin({\bf{G}}^{(s)}{\cdot}{\bf{r}})$.


\begin{thebibliography}{50}%
\makeatletter
\providecommand \@ifxundefined [1]{%
 \@ifx{#1\undefined}
}%
\providecommand \@ifnum [1]{%
 \ifnum #1\expandafter \@firstoftwo
 \else \expandafter \@secondoftwo
 \fi
}%
\providecommand \@ifx [1]{%
 \ifx #1\expandafter \@firstoftwo
 \else \expandafter \@secondoftwo
 \fi
}%
\providecommand \natexlab [1]{#1}%
\providecommand \enquote  [1]{``#1''}%
\providecommand \bibnamefont  [1]{#1}%
\providecommand \bibfnamefont [1]{#1}%
\providecommand \citenamefont [1]{#1}%
\providecommand \href@noop [0]{\@secondoftwo}%
\providecommand \href [0]{\begingroup \@sanitize@url \@href}%
\providecommand \@href[1]{\@@startlink{#1}\@@href}%
\providecommand \@@href[1]{\endgroup#1\@@endlink}%
\providecommand \@sanitize@url [0]{\catcode `\\12\catcode `\$12\catcode
  `\&12\catcode `\#12\catcode `\^12\catcode `\_12\catcode `\%12\relax}%
\providecommand \@@startlink[1]{}%
\providecommand \@@endlink[0]{}%
\providecommand \url  [0]{\begingroup\@sanitize@url \@url }%
\providecommand \@url [1]{\endgroup\@href {#1}{\urlprefix }}%
\providecommand \urlprefix  [0]{URL }%
\providecommand \Eprint [0]{\href }%
\providecommand \doibase [0]{http://dx.doi.org/}%
\providecommand \selectlanguage [0]{\@gobble}%
\providecommand \bibinfo  [0]{\@secondoftwo}%
\providecommand \bibfield  [0]{\@secondoftwo}%
\providecommand \translation [1]{[#1]}%
\providecommand \BibitemOpen [0]{}%
\providecommand \bibitemStop [0]{}%
\providecommand \bibitemNoStop [0]{.\EOS\space}%
\providecommand \EOS [0]{\spacefactor3000\relax}%
\providecommand \BibitemShut  [1]{\csname bibitem#1\endcsname}%
\let\auto@bib@innerbib\@empty
\bibitem [{\citenamefont {Cao}\ \emph {et~al.}(2018{\natexlab{a}})\citenamefont
  {Cao}, \citenamefont {Fatemi}, \citenamefont {Demir}, \citenamefont {Fang},
  \citenamefont {Tomarken}, \citenamefont {Luo}, \citenamefont
  {Sanchez-Yamagishi}, \citenamefont {Watanabe}, \citenamefont {Taniguchi},
  \citenamefont {Kaxiras}, \citenamefont {Ashoori},\ and\ \citenamefont
  {Jarillo-Herrero}}]{Cao2018}%
  \BibitemOpen
  \bibfield  {author} {\bibinfo {author} {\bibfnamefont {Y.}~\bibnamefont
  {Cao}}, \bibinfo {author} {\bibfnamefont {V.}~\bibnamefont {Fatemi}},
  \bibinfo {author} {\bibfnamefont {A.}~\bibnamefont {Demir}}, \bibinfo
  {author} {\bibfnamefont {S.}~\bibnamefont {Fang}}, \bibinfo {author}
  {\bibfnamefont {S.~L.}\ \bibnamefont {Tomarken}}, \bibinfo {author}
  {\bibfnamefont {J.~Y.}\ \bibnamefont {Luo}}, \bibinfo {author} {\bibfnamefont
  {J.~D.}\ \bibnamefont {Sanchez-Yamagishi}}, \bibinfo {author} {\bibfnamefont
  {K.}~\bibnamefont {Watanabe}}, \bibinfo {author} {\bibfnamefont
  {T.}~\bibnamefont {Taniguchi}}, \bibinfo {author} {\bibfnamefont
  {E.}~\bibnamefont {Kaxiras}}, \bibinfo {author} {\bibfnamefont {R.~C.}\
  \bibnamefont {Ashoori}}, \ and\ \bibinfo {author} {\bibfnamefont
  {P.}~\bibnamefont {Jarillo-Herrero}},\ }\bibfield  {title} {\enquote
  {\bibinfo {title} {Correlated insulator behaviour at half-filling in
  magic-angle graphene superlattices},}\ }\href {\doibase 10.1038/nature26154}
  {\bibfield  {journal} {\bibinfo  {journal} {Nature (London)}\ }\textbf
  {\bibinfo {volume} {556}},\ \bibinfo {pages} {80} (\bibinfo {year}
  {2018}{\natexlab{a}})}\BibitemShut {NoStop}%
\bibitem [{\citenamefont {Cao}\ \emph {et~al.}(2018{\natexlab{b}})\citenamefont
  {Cao}, \citenamefont {Fatemi}, \citenamefont {Fang}, \citenamefont
  {Watanabe}, \citenamefont {Taniguchi}, \citenamefont {Kaxiras},\ and\
  \citenamefont {Jarillo-Herrero}}]{cao2018unconventional}%
  \BibitemOpen
  \bibfield  {author} {\bibinfo {author} {\bibfnamefont {Y.}~\bibnamefont
  {Cao}}, \bibinfo {author} {\bibfnamefont {V.}~\bibnamefont {Fatemi}},
  \bibinfo {author} {\bibfnamefont {S.}~\bibnamefont {Fang}}, \bibinfo {author}
  {\bibfnamefont {K.}~\bibnamefont {Watanabe}}, \bibinfo {author}
  {\bibfnamefont {T.}~\bibnamefont {Taniguchi}}, \bibinfo {author}
  {\bibfnamefont {E.}~\bibnamefont {Kaxiras}}, \ and\ \bibinfo {author}
  {\bibfnamefont {P.}~\bibnamefont {Jarillo-Herrero}},\ }\bibfield  {title}
  {\enquote {\bibinfo {title} {Unconventional superconductivity in magic-angle
  graphene superlattices},}\ }\href {http://dx.doi.org/10.1038/nature26160}
  {\bibfield  {journal} {\bibinfo  {journal} {Nature (London)}\ }\textbf
  {\bibinfo {volume} {556}},\ \bibinfo {pages} {43} (\bibinfo {year}
  {2018}{\natexlab{b}})}\BibitemShut {NoStop}%
\bibitem [{\citenamefont {Sharpe}\ \emph {et~al.}(2019)\citenamefont {Sharpe},
  \citenamefont {Fox}, \citenamefont {Barnard}, \citenamefont {Finney},
  \citenamefont {Watanabe}, \citenamefont {Taniguchi}, \citenamefont
  {Kastner},\ and\ \citenamefont {Goldhaber-Gordon}}]{EmergentSharpe605}%
  \BibitemOpen
  \bibfield  {author} {\bibinfo {author} {\bibfnamefont {A.~L.}\ \bibnamefont
  {Sharpe}}, \bibinfo {author} {\bibfnamefont {E.~J.}\ \bibnamefont {Fox}},
  \bibinfo {author} {\bibfnamefont {A.~W.}\ \bibnamefont {Barnard}}, \bibinfo
  {author} {\bibfnamefont {J.}~\bibnamefont {Finney}}, \bibinfo {author}
  {\bibfnamefont {K.}~\bibnamefont {Watanabe}}, \bibinfo {author}
  {\bibfnamefont {T.}~\bibnamefont {Taniguchi}}, \bibinfo {author}
  {\bibfnamefont {M.~A.}\ \bibnamefont {Kastner}}, \ and\ \bibinfo {author}
  {\bibfnamefont {D.}~\bibnamefont {Goldhaber-Gordon}},\ }\bibfield  {title}
  {\enquote {\bibinfo {title} {Emergent ferromagnetism near three-quarters
  filling in twisted bilayer graphene},}\ }\href {\doibase
  10.1126/science.aaw3780} {\bibfield  {journal} {\bibinfo  {journal}
  {Science}\ }\textbf {\bibinfo {volume} {365}},\ \bibinfo {pages} {605}
  (\bibinfo {year} {2019})}\BibitemShut {NoStop}%
\bibitem [{\citenamefont {Lu}\ \emph {et~al.}(2019)\citenamefont {Lu},
  \citenamefont {Stepanov}, \citenamefont {Yang}, \citenamefont {Xie},
  \citenamefont {Aamir}, \citenamefont {Das}, \citenamefont {Urgell},
  \citenamefont {Watanabe}, \citenamefont {Taniguchi}, \citenamefont {Zhang},
  \citenamefont {Bachtold}, \citenamefont {MacDonald},\ and\ \citenamefont
  {Efetov}}]{lu2019superconductors}%
  \BibitemOpen
  \bibfield  {author} {\bibinfo {author} {\bibfnamefont {X.}~\bibnamefont
  {Lu}}, \bibinfo {author} {\bibfnamefont {P.}~\bibnamefont {Stepanov}},
  \bibinfo {author} {\bibfnamefont {W.}~\bibnamefont {Yang}}, \bibinfo {author}
  {\bibfnamefont {M.}~\bibnamefont {Xie}}, \bibinfo {author} {\bibfnamefont
  {M.~A.}\ \bibnamefont {Aamir}}, \bibinfo {author} {\bibfnamefont
  {I.}~\bibnamefont {Das}}, \bibinfo {author} {\bibfnamefont {C.}~\bibnamefont
  {Urgell}}, \bibinfo {author} {\bibfnamefont {K.}~\bibnamefont {Watanabe}},
  \bibinfo {author} {\bibfnamefont {T.}~\bibnamefont {Taniguchi}}, \bibinfo
  {author} {\bibfnamefont {G.}~\bibnamefont {Zhang}}, \bibinfo {author}
  {\bibfnamefont {A.}~\bibnamefont {Bachtold}}, \bibinfo {author}
  {\bibfnamefont {A.~H.}\ \bibnamefont {MacDonald}}, \ and\ \bibinfo {author}
  {\bibfnamefont {D.~K.}\ \bibnamefont {Efetov}},\ }\bibfield  {title}
  {\enquote {\bibinfo {title} {Superconductors, orbital magnets, and correlated
  states in magic angle bilayer graphene},}\ }\href {\doibase
  10.1038/s41586-019-1695-0} {\bibfield  {journal} {\bibinfo  {journal} {Nature
  (London)}\ }\textbf {\bibinfo {volume} {574}},\ \bibinfo {pages} {653}
  (\bibinfo {year} {2019})}\BibitemShut {NoStop}%
\bibitem [{\citenamefont {Bistritzer}\ and\ \citenamefont
  {MacDonald}(2011)}]{Bistritzer12233}%
  \BibitemOpen
  \bibfield  {author} {\bibinfo {author} {\bibfnamefont {R.}~\bibnamefont
  {Bistritzer}}\ and\ \bibinfo {author} {\bibfnamefont {A.~H.}\ \bibnamefont
  {MacDonald}},\ }\bibfield  {title} {\enquote {\bibinfo {title} {{M}oir{\'e}
  bands in twisted double-layer graphene},}\ }\href {\doibase
  10.1073/pnas.1108174108} {\bibfield  {journal} {\bibinfo  {journal} {Proc.
  Natl. Acad. Sci. U.S.A.}\ }\textbf {\bibinfo {volume} {108}},\ \bibinfo
  {pages} {12233} (\bibinfo {year} {2011})}\BibitemShut {NoStop}%
\bibitem [{\citenamefont {Tarnopolsky}\ \emph {et~al.}(2019)\citenamefont
  {Tarnopolsky}, \citenamefont {Kruchkov},\ and\ \citenamefont
  {Vishwanath}}]{OriginPhysRevLett.122.106405}%
  \BibitemOpen
  \bibfield  {author} {\bibinfo {author} {\bibfnamefont {G.}~\bibnamefont
  {Tarnopolsky}}, \bibinfo {author} {\bibfnamefont {A.~Jura}\ \bibnamefont
  {Kruchkov}}, \ and\ \bibinfo {author} {\bibfnamefont {A.}~\bibnamefont
  {Vishwanath}},\ }\bibfield  {title} {\enquote {\bibinfo {title} {Origin of
  {M}agic {A}ngles in {T}wisted {B}ilayer {G}raphene},}\ }\href {\doibase
  10.1103/PhysRevLett.122.106405} {\bibfield  {journal} {\bibinfo  {journal}
  {Phys. Rev. Lett.}\ }\textbf {\bibinfo {volume} {122}},\ \bibinfo {pages}
  {106405} (\bibinfo {year} {2019})}\BibitemShut {NoStop}%
\bibitem [{\citenamefont {Lopes~dos Santos}\ \emph {et~al.}(2007)\citenamefont
  {Lopes~dos Santos}, \citenamefont {Peres},\ and\ \citenamefont
  {Castro~Neto}}]{LopesdosSantos2007}%
  \BibitemOpen
  \bibfield  {author} {\bibinfo {author} {\bibfnamefont {J.~M.~B.}\
  \bibnamefont {Lopes~dos Santos}}, \bibinfo {author} {\bibfnamefont
  {N.~M.~R.}\ \bibnamefont {Peres}}, \ and\ \bibinfo {author} {\bibfnamefont
  {A.~H.}\ \bibnamefont {Castro~Neto}},\ }\bibfield  {title} {\enquote
  {\bibinfo {title} {Graphene bilayer with a twist: Electronic structure},}\
  }\href {\doibase 10.1103/PhysRevLett.99.256802} {\bibfield  {journal}
  {\bibinfo  {journal} {Phys. Rev. Lett.}\ }\textbf {\bibinfo {volume} {99}},\
  \bibinfo {pages} {256802} (\bibinfo {year} {2007})}\BibitemShut {NoStop}%
\bibitem [{\citenamefont {Su\'arez~Morell}\ \emph {et~al.}(2010)\citenamefont
  {Su\'arez~Morell}, \citenamefont {Correa}, \citenamefont {Vargas},
  \citenamefont {Pacheco},\ and\ \citenamefont {Barticevic}}]{Morell2010}%
  \BibitemOpen
  \bibfield  {author} {\bibinfo {author} {\bibfnamefont {E.}~\bibnamefont
  {Su\'arez~Morell}}, \bibinfo {author} {\bibfnamefont {J.~D.}\ \bibnamefont
  {Correa}}, \bibinfo {author} {\bibfnamefont {P.}~\bibnamefont {Vargas}},
  \bibinfo {author} {\bibfnamefont {M.}~\bibnamefont {Pacheco}}, \ and\
  \bibinfo {author} {\bibfnamefont {Z.}~\bibnamefont {Barticevic}},\ }\bibfield
   {title} {\enquote {\bibinfo {title} {Flat bands in slightly twisted bilayer
  graphene: Tight-binding calculations},}\ }\href {\doibase
  10.1103/PhysRevB.82.121407} {\bibfield  {journal} {\bibinfo  {journal} {Phys.
  Rev. B}\ }\textbf {\bibinfo {volume} {82}},\ \bibinfo {pages} {121407}
  (\bibinfo {year} {2010})}\BibitemShut {NoStop}%
\bibitem [{\citenamefont {Moon}\ and\ \citenamefont
  {Koshino}(2012)}]{Moon2012}%
  \BibitemOpen
  \bibfield  {author} {\bibinfo {author} {\bibfnamefont {P.}~\bibnamefont
  {Moon}}\ and\ \bibinfo {author} {\bibfnamefont {M.}~\bibnamefont {Koshino}},\
  }\bibfield  {title} {\enquote {\bibinfo {title} {Energy spectrum and quantum
  hall effect in twisted bilayer graphene},}\ }\href {\doibase
  10.1103/PhysRevB.85.195458} {\bibfield  {journal} {\bibinfo  {journal} {Phys.
  Rev. B}\ }\textbf {\bibinfo {volume} {85}},\ \bibinfo {pages} {195458}
  (\bibinfo {year} {2012})}\BibitemShut {NoStop}%
\bibitem [{\citenamefont {Trambly~de Laissardi\`ere}\ \emph
  {et~al.}(2012)\citenamefont {Trambly~de Laissardi\`ere}, \citenamefont
  {Mayou},\ and\ \citenamefont {Magaud}}]{Trambly2012}%
  \BibitemOpen
  \bibfield  {author} {\bibinfo {author} {\bibfnamefont {G.}~\bibnamefont
  {Trambly~de Laissardi\`ere}}, \bibinfo {author} {\bibfnamefont
  {D.}~\bibnamefont {Mayou}}, \ and\ \bibinfo {author} {\bibfnamefont
  {L.}~\bibnamefont {Magaud}},\ }\bibfield  {title} {\enquote {\bibinfo {title}
  {Numerical studies of confined states in rotated bilayers of graphene},}\
  }\href {\doibase 10.1103/PhysRevB.86.125413} {\bibfield  {journal} {\bibinfo
  {journal} {Phys. Rev. B}\ }\textbf {\bibinfo {volume} {86}},\ \bibinfo
  {pages} {125413} (\bibinfo {year} {2012})}\BibitemShut {NoStop}%
\bibitem [{\citenamefont {Lopes~dos Santos}\ \emph {et~al.}(2012)\citenamefont
  {Lopes~dos Santos}, \citenamefont {Peres},\ and\ \citenamefont
  {Castro~Neto}}]{LopesdosSantos2012}%
  \BibitemOpen
  \bibfield  {author} {\bibinfo {author} {\bibfnamefont {J.~M.~B.}\
  \bibnamefont {Lopes~dos Santos}}, \bibinfo {author} {\bibfnamefont
  {N.~M.~R.}\ \bibnamefont {Peres}}, \ and\ \bibinfo {author} {\bibfnamefont
  {A.~H.}\ \bibnamefont {Castro~Neto}},\ }\bibfield  {title} {\enquote
  {\bibinfo {title} {Continuum model of the twisted graphene bilayer},}\ }\href
  {\doibase 10.1103/PhysRevB.86.155449} {\bibfield  {journal} {\bibinfo
  {journal} {Phys. Rev. B}\ }\textbf {\bibinfo {volume} {86}},\ \bibinfo
  {pages} {155449} (\bibinfo {year} {2012})}\BibitemShut {NoStop}%
\bibitem [{\citenamefont {Fang}\ and\ \citenamefont
  {Kaxiras}(2016)}]{Fang2016}%
  \BibitemOpen
  \bibfield  {author} {\bibinfo {author} {\bibfnamefont {S.}~\bibnamefont
  {Fang}}\ and\ \bibinfo {author} {\bibfnamefont {E.}~\bibnamefont {Kaxiras}},\
  }\bibfield  {title} {\enquote {\bibinfo {title} {Electronic structure theory
  of weakly interacting bilayers},}\ }\href {\doibase
  10.1103/PhysRevB.93.235153} {\bibfield  {journal} {\bibinfo  {journal} {Phys.
  Rev. B}\ }\textbf {\bibinfo {volume} {93}},\ \bibinfo {pages} {235153}
  (\bibinfo {year} {2016})}\BibitemShut {NoStop}%
\bibitem [{\citenamefont {Yankowitz}\ \emph {et~al.}(2019)\citenamefont
  {Yankowitz}, \citenamefont {Chen}, \citenamefont {Polshyn}, \citenamefont
  {Zhang}, \citenamefont {Watanabe}, \citenamefont {Taniguchi}, \citenamefont
  {Graf}, \citenamefont {Young},\ and\ \citenamefont
  {Dean}}]{TuningYankowitz1059}%
  \BibitemOpen
  \bibfield  {author} {\bibinfo {author} {\bibfnamefont {M.}~\bibnamefont
  {Yankowitz}}, \bibinfo {author} {\bibfnamefont {S.}~\bibnamefont {Chen}},
  \bibinfo {author} {\bibfnamefont {H.}~\bibnamefont {Polshyn}}, \bibinfo
  {author} {\bibfnamefont {Y.}~\bibnamefont {Zhang}}, \bibinfo {author}
  {\bibfnamefont {K.}~\bibnamefont {Watanabe}}, \bibinfo {author}
  {\bibfnamefont {T.}~\bibnamefont {Taniguchi}}, \bibinfo {author}
  {\bibfnamefont {D.}~\bibnamefont {Graf}}, \bibinfo {author} {\bibfnamefont
  {A.~F.}\ \bibnamefont {Young}}, \ and\ \bibinfo {author} {\bibfnamefont
  {C.~R.}\ \bibnamefont {Dean}},\ }\bibfield  {title} {\enquote {\bibinfo
  {title} {Tuning superconductivity in twisted bilayer graphene},}\ }\href
  {\doibase 10.1126/science.aav1910} {\bibfield  {journal} {\bibinfo  {journal}
  {Science}\ }\textbf {\bibinfo {volume} {363}},\ \bibinfo {pages} {1059}
  (\bibinfo {year} {2019})}\BibitemShut {NoStop}%
\bibitem [{\citenamefont {Carr}\ \emph
  {et~al.}(2018{\natexlab{a}})\citenamefont {Carr}, \citenamefont {Fang},
  \citenamefont {Jarillo-Herrero},\ and\ \citenamefont
  {Kaxiras}}]{PressurePhysRevB.98.085144}%
  \BibitemOpen
  \bibfield  {author} {\bibinfo {author} {\bibfnamefont {S.}~\bibnamefont
  {Carr}}, \bibinfo {author} {\bibfnamefont {S.}~\bibnamefont {Fang}}, \bibinfo
  {author} {\bibfnamefont {P.}~\bibnamefont {Jarillo-Herrero}}, \ and\ \bibinfo
  {author} {\bibfnamefont {E.}~\bibnamefont {Kaxiras}},\ }\bibfield  {title}
  {\enquote {\bibinfo {title} {Pressure dependence of the magic twist angle in
  graphene superlattices},}\ }\href {\doibase 10.1103/PhysRevB.98.085144}
  {\bibfield  {journal} {\bibinfo  {journal} {Phys. Rev. B}\ }\textbf {\bibinfo
  {volume} {98}},\ \bibinfo {pages} {085144} (\bibinfo {year}
  {2018}{\natexlab{a}})}\BibitemShut {NoStop}%
\bibitem [{\citenamefont {Chittari}\ \emph {et~al.}(2018)\citenamefont
  {Chittari}, \citenamefont {Leconte}, \citenamefont {Javvaji},\ and\
  \citenamefont {Jung}}]{Pressurechittari2018pressure}%
  \BibitemOpen
  \bibfield  {author} {\bibinfo {author} {\bibfnamefont {B.~L.}\ \bibnamefont
  {Chittari}}, \bibinfo {author} {\bibfnamefont {N.}~\bibnamefont {Leconte}},
  \bibinfo {author} {\bibfnamefont {S.}~\bibnamefont {Javvaji}}, \ and\
  \bibinfo {author} {\bibfnamefont {J.}~\bibnamefont {Jung}},\ }\bibfield
  {title} {\enquote {\bibinfo {title} {Pressure induced compression of
  flatbands in twisted bilayer graphene},}\ }\href
  {https://iopscience.iop.org/article/10.1088/2516-1075/aaead3/meta} {\bibfield
   {journal} {\bibinfo  {journal} {Electronic Structure}\ }\textbf {\bibinfo
  {volume} {1}},\ \bibinfo {pages} {015001} (\bibinfo {year}
  {2018})}\BibitemShut {NoStop}%
\bibitem [{\citenamefont {Alden}\ \emph {et~al.}(2013)\citenamefont {Alden},
  \citenamefont {Tsen}, \citenamefont {Huang}, \citenamefont {Hovden},
  \citenamefont {Brown}, \citenamefont {Park}, \citenamefont {Muller},\ and\
  \citenamefont {McEuen}}]{McEuenBLG13}%
  \BibitemOpen
  \bibfield  {author} {\bibinfo {author} {\bibfnamefont {J.~S.}\ \bibnamefont
  {Alden}}, \bibinfo {author} {\bibfnamefont {A.~W.}\ \bibnamefont {Tsen}},
  \bibinfo {author} {\bibfnamefont {P.~Y.}\ \bibnamefont {Huang}}, \bibinfo
  {author} {\bibfnamefont {R.}~\bibnamefont {Hovden}}, \bibinfo {author}
  {\bibfnamefont {L.}~\bibnamefont {Brown}}, \bibinfo {author} {\bibfnamefont
  {J.}~\bibnamefont {Park}}, \bibinfo {author} {\bibfnamefont {D.~A.}\
  \bibnamefont {Muller}}, \ and\ \bibinfo {author} {\bibfnamefont {P.~L.}\
  \bibnamefont {McEuen}},\ }\bibfield  {title} {\enquote {\bibinfo {title}
  {Strain solitons and topological defects in bilayer graphene},}\ }\href
  {\doibase 10.1073/pnas.1309394110} {\bibfield  {journal} {\bibinfo  {journal}
  {Proc. Natl. Acad. Sci. U.S.A.}\ }\textbf {\bibinfo {volume} {110}},\
  \bibinfo {pages} {11256} (\bibinfo {year} {2013})}\BibitemShut {NoStop}%
\bibitem [{\citenamefont {Uchida}\ \emph {et~al.}(2014)\citenamefont {Uchida},
  \citenamefont {Furuya}, \citenamefont {Iwata},\ and\ \citenamefont
  {Oshiyama}}]{Uchida2014}%
  \BibitemOpen
  \bibfield  {author} {\bibinfo {author} {\bibfnamefont {K.}~\bibnamefont
  {Uchida}}, \bibinfo {author} {\bibfnamefont {S.}~\bibnamefont {Furuya}},
  \bibinfo {author} {\bibfnamefont {J.-I.}\ \bibnamefont {Iwata}}, \ and\
  \bibinfo {author} {\bibfnamefont {A.}~\bibnamefont {Oshiyama}},\ }\bibfield
  {title} {\enquote {\bibinfo {title} {Atomic corrugation and electron
  localization due to {M}oir{\'e} patterns in twisted bilayer graphenes},}\
  }\href {\doibase 10.1103/PhysRevB.90.155451} {\bibfield  {journal} {\bibinfo
  {journal} {Phys. Rev. B}\ }\textbf {\bibinfo {volume} {90}},\ \bibinfo
  {pages} {155451} (\bibinfo {year} {2014})}\BibitemShut {NoStop}%
\bibitem [{\citenamefont {van Wijk}\ \emph {et~al.}(2015)\citenamefont {van
  Wijk}, \citenamefont {Schuring}, \citenamefont {Katsnelson},\ and\
  \citenamefont {Fasolino}}]{Wijk2015}%
  \BibitemOpen
  \bibfield  {author} {\bibinfo {author} {\bibfnamefont {M.~M.}\ \bibnamefont
  {van Wijk}}, \bibinfo {author} {\bibfnamefont {A.}~\bibnamefont {Schuring}},
  \bibinfo {author} {\bibfnamefont {M.~I.}\ \bibnamefont {Katsnelson}}, \ and\
  \bibinfo {author} {\bibfnamefont {A.}~\bibnamefont {Fasolino}},\ }\bibfield
  {title} {\enquote {\bibinfo {title} {Relaxation of {M}oir{\'e} patterns for
  slightly misaligned identical lattices: graphene on graphite},}\ }\href
  {\doibase 10.1088/2053-1583/2/3/034010} {\bibfield  {journal} {\bibinfo
  {journal} {2D Mater.}\ }\textbf {\bibinfo {volume} {2}},\ \bibinfo {pages}
  {034010} (\bibinfo {year} {2015})}\BibitemShut {NoStop}%
\bibitem [{\citenamefont {Dai}\ \emph {et~al.}(2016)\citenamefont {Dai},
  \citenamefont {Xiang},\ and\ \citenamefont {Srolovitz}}]{Dai2016}%
  \BibitemOpen
  \bibfield  {author} {\bibinfo {author} {\bibfnamefont {S.}~\bibnamefont
  {Dai}}, \bibinfo {author} {\bibfnamefont {Y.}~\bibnamefont {Xiang}}, \ and\
  \bibinfo {author} {\bibfnamefont {D.~J.}\ \bibnamefont {Srolovitz}},\
  }\bibfield  {title} {\enquote {\bibinfo {title} {Twisted bilayer graphene:
  {M}oir{\'e} with a twist},}\ }\href {\doibase 10.1021/acs.nanolett.6b02870}
  {\bibfield  {journal} {\bibinfo  {journal} {Nano Lett.}\ }\textbf {\bibinfo
  {volume} {16}},\ \bibinfo {pages} {5923} (\bibinfo {year}
  {2016})}\BibitemShut {NoStop}%
\bibitem [{\citenamefont {Jain}\ \emph {et~al.}(2017)\citenamefont {Jain},
  \citenamefont {Juri{\v{c}}i{\'c}},\ and\ \citenamefont {Barkema}}]{Jain2017}%
  \BibitemOpen
  \bibfield  {author} {\bibinfo {author} {\bibfnamefont {S.~K.}\ \bibnamefont
  {Jain}}, \bibinfo {author} {\bibfnamefont {V.}~\bibnamefont
  {Juri{\v{c}}i{\'c}}}, \ and\ \bibinfo {author} {\bibfnamefont {G.~T.}\
  \bibnamefont {Barkema}},\ }\bibfield  {title} {\enquote {\bibinfo {title}
  {Structure of twisted and buckled bilayer graphene},}\ }\href {\doibase
  10.1088/2053-1583/4/1/015018} {\bibfield  {journal} {\bibinfo  {journal} {2D
  Mater.}\ }\textbf {\bibinfo {volume} {4}},\ \bibinfo {pages} {015018}
  (\bibinfo {year} {2017})}\BibitemShut {NoStop}%
\bibitem [{\citenamefont {Nam}\ and\ \citenamefont {Koshino}(2017)}]{Nam2017}%
  \BibitemOpen
  \bibfield  {author} {\bibinfo {author} {\bibfnamefont {N.~N.~T.}\
  \bibnamefont {Nam}}\ and\ \bibinfo {author} {\bibfnamefont {M.}~\bibnamefont
  {Koshino}},\ }\bibfield  {title} {\enquote {\bibinfo {title} {Lattice
  relaxation and energy band modulation in twisted bilayer graphene},}\ }\href
  {\doibase 10.1103/PhysRevB.96.075311} {\bibfield  {journal} {\bibinfo
  {journal} {Phys. Rev. B}\ }\textbf {\bibinfo {volume} {96}},\ \bibinfo
  {pages} {075311} (\bibinfo {year} {2017})}\BibitemShut {NoStop}%
\bibitem [{\citenamefont {Gargiulo}\ and\ \citenamefont
  {Yazyev}(2018)}]{Gargiulo2018}%
  \BibitemOpen
  \bibfield  {author} {\bibinfo {author} {\bibfnamefont {F.}~\bibnamefont
  {Gargiulo}}\ and\ \bibinfo {author} {\bibfnamefont {O.~V.}\ \bibnamefont
  {Yazyev}},\ }\bibfield  {title} {\enquote {\bibinfo {title} {Structural and
  electronic transformation in low-angle twisted bilayer graphene},}\ }\href
  {\doibase 10.1088/2053-1583/aa9640} {\bibfield  {journal} {\bibinfo
  {journal} {2D Mater.}\ }\textbf {\bibinfo {volume} {5}},\ \bibinfo {pages}
  {015019} (\bibinfo {year} {2018})}\BibitemShut {NoStop}%
\bibitem [{\citenamefont {Carr}\ \emph
  {et~al.}(2018{\natexlab{b}})\citenamefont {Carr}, \citenamefont {Massatt},
  \citenamefont {Torrisi}, \citenamefont {Cazeaux}, \citenamefont {Luskin},\
  and\ \citenamefont {Kaxiras}}]{carr2018relaxation}%
  \BibitemOpen
  \bibfield  {author} {\bibinfo {author} {\bibfnamefont {S.}~\bibnamefont
  {Carr}}, \bibinfo {author} {\bibfnamefont {D.}~\bibnamefont {Massatt}},
  \bibinfo {author} {\bibfnamefont {S.~B.}\ \bibnamefont {Torrisi}}, \bibinfo
  {author} {\bibfnamefont {P.}~\bibnamefont {Cazeaux}}, \bibinfo {author}
  {\bibfnamefont {M.}~\bibnamefont {Luskin}}, \ and\ \bibinfo {author}
  {\bibfnamefont {E.}~\bibnamefont {Kaxiras}},\ }\bibfield  {title} {\enquote
  {\bibinfo {title} {Relaxation and domain formation in incommensurate
  two-dimensional heterostructures},}\ }\href {\doibase
  10.1103/PhysRevB.98.224102} {\bibfield  {journal} {\bibinfo  {journal} {Phys.
  Rev. B}\ }\textbf {\bibinfo {volume} {98}},\ \bibinfo {pages} {224102}
  (\bibinfo {year} {2018}{\natexlab{b}})}\BibitemShut {NoStop}%
\bibitem [{\citenamefont {Lin}\ \emph {et~al.}(2018)\citenamefont {Lin},
  \citenamefont {Liu},\ and\ \citenamefont
  {Tom\'anek}}]{ShearPhysRevB.98.195432}%
  \BibitemOpen
  \bibfield  {author} {\bibinfo {author} {\bibfnamefont {X.}~\bibnamefont
  {Lin}}, \bibinfo {author} {\bibfnamefont {D.}~\bibnamefont {Liu}}, \ and\
  \bibinfo {author} {\bibfnamefont {D.}~\bibnamefont {Tom\'anek}},\ }\bibfield
  {title} {\enquote {\bibinfo {title} {Shear instability in twisted bilayer
  graphene},}\ }\href {\doibase 10.1103/PhysRevB.98.195432} {\bibfield
  {journal} {\bibinfo  {journal} {Phys. Rev. B}\ }\textbf {\bibinfo {volume}
  {98}},\ \bibinfo {pages} {195432} (\bibinfo {year} {2018})}\BibitemShut
  {NoStop}%
\bibitem [{\citenamefont {Yoo}\ \emph {et~al.}(2019)\citenamefont {Yoo},
  \citenamefont {Engelke}, \citenamefont {Carr}, \citenamefont {Fang},
  \citenamefont {Zhang}, \citenamefont {Cazeaux}, \citenamefont {Sung},
  \citenamefont {Hovden}, \citenamefont {Tsen}, \citenamefont {Taniguchi},
  \citenamefont {Watanabe}, \citenamefont {Yi}, \citenamefont {Kim},
  \citenamefont {Luskin}, \citenamefont {Tadmor}, \citenamefont {Kaxiras},\
  and\ \citenamefont {Kim}}]{Atomicyoo2019atomic}%
  \BibitemOpen
  \bibfield  {author} {\bibinfo {author} {\bibfnamefont {H.}~\bibnamefont
  {Yoo}}, \bibinfo {author} {\bibfnamefont {R.}~\bibnamefont {Engelke}},
  \bibinfo {author} {\bibfnamefont {S.}~\bibnamefont {Carr}}, \bibinfo {author}
  {\bibfnamefont {S.}~\bibnamefont {Fang}}, \bibinfo {author} {\bibfnamefont
  {K.}~\bibnamefont {Zhang}}, \bibinfo {author} {\bibfnamefont
  {P.}~\bibnamefont {Cazeaux}}, \bibinfo {author} {\bibfnamefont {S.~H.}\
  \bibnamefont {Sung}}, \bibinfo {author} {\bibfnamefont {R.}~\bibnamefont
  {Hovden}}, \bibinfo {author} {\bibfnamefont {A.~W.}\ \bibnamefont {Tsen}},
  \bibinfo {author} {\bibfnamefont {T.}~\bibnamefont {Taniguchi}}, \bibinfo
  {author} {\bibfnamefont {K.}~\bibnamefont {Watanabe}}, \bibinfo {author}
  {\bibfnamefont {G.-C.}\ \bibnamefont {Yi}}, \bibinfo {author} {\bibfnamefont
  {M.}~\bibnamefont {Kim}}, \bibinfo {author} {\bibfnamefont {M.}~\bibnamefont
  {Luskin}}, \bibinfo {author} {\bibfnamefont {E.~B.}\ \bibnamefont {Tadmor}},
  \bibinfo {author} {\bibfnamefont {E.}~\bibnamefont {Kaxiras}}, \ and\
  \bibinfo {author} {\bibfnamefont {P.}~\bibnamefont {Kim}},\ }\bibfield
  {title} {\enquote {\bibinfo {title} {Atomic and electronic reconstruction at
  the van der waals interface in twisted bilayer graphene},}\ }\href {\doibase
  10.1038/s41563-019-0346-z} {\bibfield  {journal} {\bibinfo  {journal} {Nat.
  Mater.}\ }\textbf {\bibinfo {volume} {18}},\ \bibinfo {pages} {448} (\bibinfo
  {year} {2019})}\BibitemShut {NoStop}%
\bibitem [{\citenamefont {Lucignano}\ \emph {et~al.}(2019)\citenamefont
  {Lucignano}, \citenamefont {Alf\`e}, \citenamefont {Cataudella},
  \citenamefont {Ninno},\ and\ \citenamefont
  {Cantele}}]{CrucialPhysRevB.99.195419}%
  \BibitemOpen
  \bibfield  {author} {\bibinfo {author} {\bibfnamefont {P.}~\bibnamefont
  {Lucignano}}, \bibinfo {author} {\bibfnamefont {D.}~\bibnamefont {Alf\`e}},
  \bibinfo {author} {\bibfnamefont {V.}~\bibnamefont {Cataudella}}, \bibinfo
  {author} {\bibfnamefont {D.}~\bibnamefont {Ninno}}, \ and\ \bibinfo {author}
  {\bibfnamefont {G.}~\bibnamefont {Cantele}},\ }\bibfield  {title} {\enquote
  {\bibinfo {title} {Crucial role of atomic corrugation on the flat bands and
  energy gaps of twisted bilayer graphene at the magic angle
  $\ensuremath{\theta}\ensuremath{\sim}1.{08}^{\ensuremath{\circ}}$},}\ }\href
  {\doibase 10.1103/PhysRevB.99.195419} {\bibfield  {journal} {\bibinfo
  {journal} {Phys. Rev. B}\ }\textbf {\bibinfo {volume} {99}},\ \bibinfo
  {pages} {195419} (\bibinfo {year} {2019})}\BibitemShut {NoStop}%
\bibitem [{\citenamefont {Guinea}\ and\ \citenamefont
  {Walet}(2019)}]{ContinuumPhysRevB.99.205134}%
  \BibitemOpen
  \bibfield  {author} {\bibinfo {author} {\bibfnamefont {F.}~\bibnamefont
  {Guinea}}\ and\ \bibinfo {author} {\bibfnamefont {N.~R.}\ \bibnamefont
  {Walet}},\ }\bibfield  {title} {\enquote {\bibinfo {title} {Continuum models
  for twisted bilayer graphene: {E}ffect of lattice deformation and hopping
  parameters},}\ }\href {\doibase 10.1103/PhysRevB.99.205134} {\bibfield
  {journal} {\bibinfo  {journal} {Phys. Rev. B}\ }\textbf {\bibinfo {volume}
  {99}},\ \bibinfo {pages} {205134} (\bibinfo {year} {2019})}\BibitemShut
  {NoStop}%
\bibitem [{\citenamefont {Shen}\ \emph {et~al.}()\citenamefont {Shen},
  \citenamefont {Li}, \citenamefont {Wang}, \citenamefont {Zhao}, \citenamefont
  {Tang}, \citenamefont {Liu}, \citenamefont {Tian}, \citenamefont {Chu},
  \citenamefont {Watanabe}, \citenamefont {Taniguchi}, \citenamefont {Yang},
  \citenamefont {Meng}, \citenamefont {Shi},\ and\ \citenamefont
  {Zhang}}]{Observation2019}%
  \BibitemOpen
  \bibfield  {author} {\bibinfo {author} {\bibfnamefont {C.}~\bibnamefont
  {Shen}}, \bibinfo {author} {\bibfnamefont {N.}~\bibnamefont {Li}}, \bibinfo
  {author} {\bibfnamefont {S.}~\bibnamefont {Wang}}, \bibinfo {author}
  {\bibfnamefont {Y.}~\bibnamefont {Zhao}}, \bibinfo {author} {\bibfnamefont
  {J.}~\bibnamefont {Tang}}, \bibinfo {author} {\bibfnamefont {J.}~\bibnamefont
  {Liu}}, \bibinfo {author} {\bibfnamefont {J.}~\bibnamefont {Tian}}, \bibinfo
  {author} {\bibfnamefont {Y.}~\bibnamefont {Chu}}, \bibinfo {author}
  {\bibfnamefont {K.}~\bibnamefont {Watanabe}}, \bibinfo {author}
  {\bibfnamefont {T.}~\bibnamefont {Taniguchi}}, \bibinfo {author}
  {\bibfnamefont {R.}~\bibnamefont {Yang}}, \bibinfo {author} {\bibfnamefont
  {Z.~Y.}\ \bibnamefont {Meng}}, \bibinfo {author} {\bibfnamefont
  {D.}~\bibnamefont {Shi}}, \ and\ \bibinfo {author} {\bibfnamefont
  {G.}~\bibnamefont {Zhang}},\ }\bibfield  {title} {\enquote {\bibinfo {title}
  {Observation of superconductivity with {T}$_c$ onset at 12{K} in electrically
  tunable twisted double bilayer graphene},}\ }\href
  {https://arxiv.org/abs/1903.06952} {\bibinfo  {journal} {arXiv:1903.06952}\
  }\BibitemShut {NoStop}%
\bibitem [{\citenamefont {Liu}\ \emph {et~al.}()\citenamefont {Liu},
  \citenamefont {Hao}, \citenamefont {Khalaf}, \citenamefont {Lee},
  \citenamefont {Watanabe}, \citenamefont {Taniguchi}, \citenamefont
  {Vishwanath},\ and\ \citenamefont {Kim}}]{Spin-polarizedliu2019}%
  \BibitemOpen
\bibfield  {journal} {  }\bibfield  {author} {\bibinfo {author} {\bibfnamefont
  {X.}~\bibnamefont {Liu}}, \bibinfo {author} {\bibfnamefont {Z.}~\bibnamefont
  {Hao}}, \bibinfo {author} {\bibfnamefont {E.}~\bibnamefont {Khalaf}},
  \bibinfo {author} {\bibfnamefont {J.~Y.}\ \bibnamefont {Lee}}, \bibinfo
  {author} {\bibfnamefont {K.}~\bibnamefont {Watanabe}}, \bibinfo {author}
  {\bibfnamefont {T.}~\bibnamefont {Taniguchi}}, \bibinfo {author}
  {\bibfnamefont {A.}~\bibnamefont {Vishwanath}}, \ and\ \bibinfo {author}
  {\bibfnamefont {P.}~\bibnamefont {Kim}},\ }\bibfield  {title} {\enquote
  {\bibinfo {title} {Spin-polarized {C}orrelated {I}nsulator and
  {S}uperconductor in {T}wisted {D}ouble {B}ilayer {G}raphene},}\ }\href
  {https://arxiv.org/abs/1903.08130} {\bibinfo  {journal} {arXiv:1903.08130}\
  }\BibitemShut {NoStop}%
\bibitem [{\citenamefont {Cao}\ \emph {et~al.}()\citenamefont {Cao},
  \citenamefont {Rodan-Legrain}, \citenamefont {Rubies-Bigord{\`a}},
  \citenamefont {Park}, \citenamefont {Watanabe}, \citenamefont {Taniguchi},\
  and\ \citenamefont {Jarillo-Herrero}}]{Electriccao2019}%
  \BibitemOpen
\bibfield  {journal} {  }\bibfield  {author} {\bibinfo {author} {\bibfnamefont
  {Y.}~\bibnamefont {Cao}}, \bibinfo {author} {\bibfnamefont {D.}~\bibnamefont
  {Rodan-Legrain}}, \bibinfo {author} {\bibfnamefont {O.}~\bibnamefont
  {Rubies-Bigord{\`a}}}, \bibinfo {author} {\bibfnamefont {J.~M.}\ \bibnamefont
  {Park}}, \bibinfo {author} {\bibfnamefont {K.}~\bibnamefont {Watanabe}},
  \bibinfo {author} {\bibfnamefont {T.}~\bibnamefont {Taniguchi}}, \ and\
  \bibinfo {author} {\bibfnamefont {P.}~\bibnamefont {Jarillo-Herrero}},\
  }\bibfield  {title} {\enquote {\bibinfo {title} {Electric {F}ield {T}unable
  {C}orrelated {S}tates and {M}agnetic {P}hase {T}ransitions in {T}wisted
  {B}ilayer-{B}ilayer {G}raphene},}\ }\href {https://arxiv.org/abs/1903.08596}
  {\bibinfo  {journal} {arXiv:1903.08596}\ }\BibitemShut {NoStop}%
\bibitem [{\citenamefont {Koshino}(2019)}]{BandPhysRevB.99.235406}%
  \BibitemOpen
\bibfield  {journal} {  }\bibfield  {author} {\bibinfo {author} {\bibfnamefont
  {M.}~\bibnamefont {Koshino}},\ }\bibfield  {title} {\enquote {\bibinfo
  {title} {Band structure and topological properties of twisted double bilayer
  graphene},}\ }\href {\doibase 10.1103/PhysRevB.99.235406} {\bibfield
  {journal} {\bibinfo  {journal} {Phys. Rev. B}\ }\textbf {\bibinfo {volume}
  {99}},\ \bibinfo {pages} {235406} (\bibinfo {year} {2019})}\BibitemShut
  {NoStop}%
\bibitem [{\citenamefont {Chebrolu}\ \emph {et~al.}(2019)\citenamefont
  {Chebrolu}, \citenamefont {Chittari},\ and\ \citenamefont
  {Jung}}]{FlatPhysRevB.99.235417}%
  \BibitemOpen
  \bibfield  {author} {\bibinfo {author} {\bibfnamefont {N.~R.}\ \bibnamefont
  {Chebrolu}}, \bibinfo {author} {\bibfnamefont {B.~L.}\ \bibnamefont
  {Chittari}}, \ and\ \bibinfo {author} {\bibfnamefont {J.}~\bibnamefont
  {Jung}},\ }\bibfield  {title} {\enquote {\bibinfo {title} {Flat bands in
  twisted double bilayer graphene},}\ }\href {\doibase
  10.1103/PhysRevB.99.235417} {\bibfield  {journal} {\bibinfo  {journal} {Phys.
  Rev. B}\ }\textbf {\bibinfo {volume} {99}},\ \bibinfo {pages} {235417}
  (\bibinfo {year} {2019})}\BibitemShut {NoStop}%
\bibitem [{\citenamefont {Choi}\ and\ \citenamefont
  {Choi}(2019)}]{IntrinsicPhysRevB.100.201402}%
  \BibitemOpen
  \bibfield  {author} {\bibinfo {author} {\bibfnamefont {Y.~W.}\ \bibnamefont
  {Choi}}\ and\ \bibinfo {author} {\bibfnamefont {H.~J.}\ \bibnamefont
  {Choi}},\ }\bibfield  {title} {\enquote {\bibinfo {title} {Intrinsic band gap
  and electrically tunable flat bands in twisted double bilayer graphene},}\
  }\href {\doibase 10.1103/PhysRevB.100.201402} {\bibfield  {journal} {\bibinfo
   {journal} {Phys. Rev. B}\ }\textbf {\bibinfo {volume} {100}},\ \bibinfo
  {pages} {201402} (\bibinfo {year} {2019})}\BibitemShut {NoStop}%
\bibitem [{\citenamefont {Liu}\ \emph {et~al.}(2019)\citenamefont {Liu},
  \citenamefont {Ma}, \citenamefont {Gao},\ and\ \citenamefont
  {Dai}}]{QuantumPhysRevX.9.031021}%
  \BibitemOpen
  \bibfield  {author} {\bibinfo {author} {\bibfnamefont {J.}~\bibnamefont
  {Liu}}, \bibinfo {author} {\bibfnamefont {Z.}~\bibnamefont {Ma}}, \bibinfo
  {author} {\bibfnamefont {J.}~\bibnamefont {Gao}}, \ and\ \bibinfo {author}
  {\bibfnamefont {X.}~\bibnamefont {Dai}},\ }\bibfield  {title} {\enquote
  {\bibinfo {title} {Quantum valley hall effect, orbital magnetism, and
  anomalous hall effect in twisted multilayer graphene systems},}\ }\href
  {\doibase 10.1103/PhysRevX.9.031021} {\bibfield  {journal} {\bibinfo
  {journal} {Phys. Rev. X}\ }\textbf {\bibinfo {volume} {9}},\ \bibinfo {pages}
  {031021} (\bibinfo {year} {2019})}\BibitemShut {NoStop}%
\bibitem [{\citenamefont {Lee}\ \emph {et~al.}(2019)\citenamefont {Lee},
  \citenamefont {Khalaf}, \citenamefont {Liu}, \citenamefont {Liu},
  \citenamefont {Hao}, \citenamefont {Kim},\ and\ \citenamefont
  {Vishwanath}}]{Theorylee2019theory}%
  \BibitemOpen
  \bibfield  {author} {\bibinfo {author} {\bibfnamefont {J.~Y.}\ \bibnamefont
  {Lee}}, \bibinfo {author} {\bibfnamefont {E.}~\bibnamefont {Khalaf}},
  \bibinfo {author} {\bibfnamefont {S.}~\bibnamefont {Liu}}, \bibinfo {author}
  {\bibfnamefont {X.}~\bibnamefont {Liu}}, \bibinfo {author} {\bibfnamefont
  {Z.}~\bibnamefont {Hao}}, \bibinfo {author} {\bibfnamefont {P.}~\bibnamefont
  {Kim}}, \ and\ \bibinfo {author} {\bibfnamefont {A.}~\bibnamefont
  {Vishwanath}},\ }\bibfield  {title} {\enquote {\bibinfo {title} {Theory of
  correlated insulating behaviour and spin-triplet superconductivity in twisted
  double bilayer graphene},}\ }\href {\doibase 10.1038/s41467-019-12981-1}
  {\bibfield  {journal} {\bibinfo  {journal} {Nat. Commun.}\ }\textbf {\bibinfo
  {volume} {10}},\ \bibinfo {pages} {5333} (\bibinfo {year}
  {2019})}\BibitemShut {NoStop}%
\bibitem [{\citenamefont {Zhou}\ \emph {et~al.}(2015)\citenamefont {Zhou},
  \citenamefont {Han}, \citenamefont {Dai}, \citenamefont {Sun},\ and\
  \citenamefont {Srolovitz}}]{Zhou2015}%
  \BibitemOpen
  \bibfield  {author} {\bibinfo {author} {\bibfnamefont {S.}~\bibnamefont
  {Zhou}}, \bibinfo {author} {\bibfnamefont {J.}~\bibnamefont {Han}}, \bibinfo
  {author} {\bibfnamefont {S.}~\bibnamefont {Dai}}, \bibinfo {author}
  {\bibfnamefont {J.}~\bibnamefont {Sun}}, \ and\ \bibinfo {author}
  {\bibfnamefont {D.~J.}\ \bibnamefont {Srolovitz}},\ }\bibfield  {title}
  {\enquote {\bibinfo {title} {{V}an der {W}aals bilayer energetics:
  Generalized stacking-fault energy of graphene, boron nitride, and
  graphene/boron nitride bilayers},}\ }\href {\doibase
  10.1103/PhysRevB.92.155438} {\bibfield  {journal} {\bibinfo  {journal} {Phys.
  Rev. B}\ }\textbf {\bibinfo {volume} {92}},\ \bibinfo {pages} {155438}
  (\bibinfo {year} {2015})}\BibitemShut {NoStop}%
\bibitem [{\citenamefont {de~Andres}\ \emph {et~al.}(2012)\citenamefont
  {de~Andres}, \citenamefont {Guinea},\ and\ \citenamefont
  {Katsnelson}}]{Andres2012}%
  \BibitemOpen
  \bibfield  {author} {\bibinfo {author} {\bibfnamefont {P.~L.}\ \bibnamefont
  {de~Andres}}, \bibinfo {author} {\bibfnamefont {F.}~\bibnamefont {Guinea}}, \
  and\ \bibinfo {author} {\bibfnamefont {M.~I.}\ \bibnamefont {Katsnelson}},\
  }\bibfield  {title} {\enquote {\bibinfo {title} {Bending modes, anharmonic
  effects, and thermal expansion coefficient in single-layer and multilayer
  graphene},}\ }\href {\doibase 10.1103/PhysRevB.86.144103} {\bibfield
  {journal} {\bibinfo  {journal} {Phys. Rev. B}\ }\textbf {\bibinfo {volume}
  {86}},\ \bibinfo {pages} {144103} (\bibinfo {year} {2012})}\BibitemShut
  {NoStop}%
\bibitem [{\citenamefont {Yankowitz}\ \emph {et~al.}(2018)\citenamefont
  {Yankowitz}, \citenamefont {Jung}, \citenamefont {Laksono}, \citenamefont
  {Leconte}, \citenamefont {Chittari}, \citenamefont {Watanabe}, \citenamefont
  {Taniguchi}, \citenamefont {Adam}, \citenamefont {Graf},\ and\ \citenamefont
  {Dean}}]{Dynamic2018May}%
  \BibitemOpen
  \bibfield  {author} {\bibinfo {author} {\bibfnamefont {M.}~\bibnamefont
  {Yankowitz}}, \bibinfo {author} {\bibfnamefont {J.}~\bibnamefont {Jung}},
  \bibinfo {author} {\bibfnamefont {E.}~\bibnamefont {Laksono}}, \bibinfo
  {author} {\bibfnamefont {N.}~\bibnamefont {Leconte}}, \bibinfo {author}
  {\bibfnamefont {B.~L.}\ \bibnamefont {Chittari}}, \bibinfo {author}
  {\bibfnamefont {K.}~\bibnamefont {Watanabe}}, \bibinfo {author}
  {\bibfnamefont {T.}~\bibnamefont {Taniguchi}}, \bibinfo {author}
  {\bibfnamefont {S.}~\bibnamefont {Adam}}, \bibinfo {author} {\bibfnamefont
  {D.}~\bibnamefont {Graf}}, \ and\ \bibinfo {author} {\bibfnamefont {C.~R.}\
  \bibnamefont {Dean}},\ }\bibfield  {title} {\enquote {\bibinfo {title}
  {Dynamic band-structure tuning of graphene moir\'{e} superlattices with
  pressure},}\ }\href {\doibase 10.1038/s41586-018-0107-1} {\bibfield
  {journal} {\bibinfo  {journal} {Nature (London)}\ }\textbf {\bibinfo {volume}
  {557}},\ \bibinfo {pages} {404} (\bibinfo {year} {2018})}\BibitemShut
  {NoStop}%
\bibitem [{\citenamefont {Lin}\ and\ \citenamefont
  {Tom{\'a}nek}(2018)}]{DT268}%
  \BibitemOpen
  \bibfield  {author} {\bibinfo {author} {\bibfnamefont {X.}~\bibnamefont
  {Lin}}\ and\ \bibinfo {author} {\bibfnamefont {D.}~\bibnamefont
  {Tom{\'a}nek}},\ }\bibfield  {title} {\enquote {\bibinfo {title} {Minimum
  model for the electronic structure of twisted bilayer graphene and related
  structures},}\ }\href {\doibase 10.1103/PhysRevB.98.081410} {\bibfield
  {journal} {\bibinfo  {journal} {Phys. Rev. B}\ }\textbf {\bibinfo {volume}
  {98}},\ \bibinfo {pages} {081410} (\bibinfo {year} {2018})}\BibitemShut
  {NoStop}%
\bibitem [{\citenamefont {Castro~Neto}\ \emph {et~al.}(2009)\citenamefont
  {Castro~Neto}, \citenamefont {Guinea}, \citenamefont {Peres}, \citenamefont
  {Novoselov},\ and\ \citenamefont {Geim}}]{CastroNeto2009}%
  \BibitemOpen
  \bibfield  {author} {\bibinfo {author} {\bibfnamefont {A.~H.}\ \bibnamefont
  {Castro~Neto}}, \bibinfo {author} {\bibfnamefont {F.}~\bibnamefont {Guinea}},
  \bibinfo {author} {\bibfnamefont {N.~M.~R.}\ \bibnamefont {Peres}}, \bibinfo
  {author} {\bibfnamefont {K.~S.}\ \bibnamefont {Novoselov}}, \ and\ \bibinfo
  {author} {\bibfnamefont {A.~K.}\ \bibnamefont {Geim}},\ }\bibfield  {title}
  {\enquote {\bibinfo {title} {The electronic properties of graphene},}\ }\href
  {\doibase 10.1103/RevModPhys.81.109} {\bibfield  {journal} {\bibinfo
  {journal} {Rev. Mod. Phys.}\ }\textbf {\bibinfo {volume} {81}},\ \bibinfo
  {pages} {109--162} (\bibinfo {year} {2009})}\BibitemShut {NoStop}%
\bibitem [{pre()}]{press-SM}%
  \BibitemOpen
  \href@noop {} {}\bibinfo {note} {See Supplemental Material for varying band
  structures with pressure ($P$) for TBG with $\theta = 1.248^{\circ}$ (Video
  1) and for TDBG with $\theta$ = 1.538$^{\circ}$ (Video 2). In Video 1, the
  data of relaxed TBG are represented by solid lines and the dashed lines
  denote rigid superlattices with interlayer distances the same as the average
  ones ($h_0$) of the relaxed structures. In Video 2, the left and right panels
  show the band structures of relaxed and rigid TDBGs, respectively, and the
  blue and red lines represent bands in the $\xi = +$ and $\xi = -$ valleys,
  respectively.}\BibitemShut {Stop}%
\bibitem [{\citenamefont {Xiao}\ \emph {et~al.}(2010)\citenamefont {Xiao},
  \citenamefont {Chang},\ and\ \citenamefont {Niu}}]{RevModPhys.82.1959}%
  \BibitemOpen
  \bibfield  {author} {\bibinfo {author} {\bibfnamefont {D.}~\bibnamefont
  {Xiao}}, \bibinfo {author} {\bibfnamefont {M.-C.}\ \bibnamefont {Chang}}, \
  and\ \bibinfo {author} {\bibfnamefont {Q.}~\bibnamefont {Niu}},\ }\bibfield
  {title} {\enquote {\bibinfo {title} {Berry phase effects on electronic
  properties},}\ }\href {\doibase 10.1103/RevModPhys.82.1959} {\bibfield
  {journal} {\bibinfo  {journal} {Rev. Mod. Phys.}\ }\textbf {\bibinfo {volume}
  {82}},\ \bibinfo {pages} {1959} (\bibinfo {year} {2010})}\BibitemShut
  {NoStop}%
\bibitem [{\citenamefont {Kresse}\ and\ \citenamefont
  {Furthm\"{u}ller}(1996{\natexlab{a}})}]{VASP1}%
  \BibitemOpen
  \bibfield  {author} {\bibinfo {author} {\bibfnamefont {G.}~\bibnamefont
  {Kresse}}\ and\ \bibinfo {author} {\bibfnamefont {J.}~\bibnamefont
  {Furthm\"{u}ller}},\ }\bibfield  {title} {\enquote {\bibinfo {title}
  {Efficient iterative schemes for ab initio total-energy calculations using a
  plane-wave basis set},}\ }\href {\doibase 10.1103/PhysRevB.54.11169}
  {\bibfield  {journal} {\bibinfo  {journal} {Phys. Rev. B}\ }\textbf {\bibinfo
  {volume} {54}},\ \bibinfo {pages} {11169} (\bibinfo {year}
  {1996}{\natexlab{a}})}\BibitemShut {NoStop}%
\bibitem [{\citenamefont {Kresse}\ and\ \citenamefont
  {Furthm\"{u}ller}(1996{\natexlab{b}})}]{VASP2}%
  \BibitemOpen
  \bibfield  {author} {\bibinfo {author} {\bibfnamefont {G.}~\bibnamefont
  {Kresse}}\ and\ \bibinfo {author} {\bibfnamefont {J.}~\bibnamefont
  {Furthm\"{u}ller}},\ }\bibfield  {title} {\enquote {\bibinfo {title}
  {Efficiency of ab-initio total energy calculations for metals and
  semiconductors using a plane-wave basis set},}\ }\href {\doibase
  https://doi.org/10.1016/0927-0256(96)00008-0} {\bibfield  {journal} {\bibinfo
   {journal} {Computat. Mater. Sci.}\ }\textbf {\bibinfo {volume} {6}},\
  \bibinfo {pages} {15} (\bibinfo {year} {1996}{\natexlab{b}})}\BibitemShut
  {NoStop}%
\bibitem [{\citenamefont {Kresse}\ and\ \citenamefont {Hafner}(1994)}]{VASP3}%
  \BibitemOpen
  \bibfield  {author} {\bibinfo {author} {\bibfnamefont {G.}~\bibnamefont
  {Kresse}}\ and\ \bibinfo {author} {\bibfnamefont {J.}~\bibnamefont
  {Hafner}},\ }\bibfield  {title} {\enquote {\bibinfo {title} {\textit{Ab
  initio} molecular-dynamics simulation of the
  liquid-metal--amorphous-semiconductor transition in germanium},}\ }\href
  {\doibase 10.1103/PhysRevB.49.14251} {\bibfield  {journal} {\bibinfo
  {journal} {Phys. Rev. B}\ }\textbf {\bibinfo {volume} {49}},\ \bibinfo
  {pages} {14251} (\bibinfo {year} {1994})}\BibitemShut {NoStop}%
\bibitem [{\citenamefont {Bl\"ochl}(1994)}]{PAW1}%
  \BibitemOpen
  \bibfield  {author} {\bibinfo {author} {\bibfnamefont {P.~E.}\ \bibnamefont
  {Bl\"ochl}},\ }\bibfield  {title} {\enquote {\bibinfo {title} {Projector
  augmented-wave method},}\ }\href {\doibase 10.1103/PhysRevB.50.17953}
  {\bibfield  {journal} {\bibinfo  {journal} {Phys. Rev. B}\ }\textbf {\bibinfo
  {volume} {50}},\ \bibinfo {pages} {17953} (\bibinfo {year}
  {1994})}\BibitemShut {NoStop}%
\bibitem [{\citenamefont {Kresse}\ and\ \citenamefont {Joubert}(1999)}]{PAW2}%
  \BibitemOpen
  \bibfield  {author} {\bibinfo {author} {\bibfnamefont {G.}~\bibnamefont
  {Kresse}}\ and\ \bibinfo {author} {\bibfnamefont {D.}~\bibnamefont
  {Joubert}},\ }\bibfield  {title} {\enquote {\bibinfo {title} {From ultrasoft
  pseudopotentials to the projector augmented-wave method},}\ }\href {\doibase
  10.1103/PhysRevB.59.1758} {\bibfield  {journal} {\bibinfo  {journal} {Phys.
  Rev. B}\ }\textbf {\bibinfo {volume} {59}},\ \bibinfo {pages} {1758}
  (\bibinfo {year} {1999})}\BibitemShut {NoStop}%
\bibitem [{\citenamefont {Klime\ifmmode~\check{s}\else \v{s}\fi{}}\ \emph
  {et~al.}(2011)\citenamefont {Klime\ifmmode~\check{s}\else \v{s}\fi{}},
  \citenamefont {Bowler},\ and\ \citenamefont {Michaelides}}]{Klime2011}%
  \BibitemOpen
  \bibfield  {author} {\bibinfo {author} {\bibfnamefont
  {Ji\ifmmode\check{r}\else\v{r}\fi{}\'{\i}}\ \bibnamefont
  {Klime\ifmmode~\check{s}\else \v{s}\fi{}}}, \bibinfo {author} {\bibfnamefont
  {D.~R.}\ \bibnamefont {Bowler}}, \ and\ \bibinfo {author} {\bibfnamefont
  {A.}~\bibnamefont {Michaelides}},\ }\bibfield  {title} {\enquote {\bibinfo
  {title} {Van der {W}aals density functionals applied to solids},}\ }\href
  {\doibase 10.1103/PhysRevB.83.195131} {\bibfield  {journal} {\bibinfo
  {journal} {Phys. Rev. B}\ }\textbf {\bibinfo {volume} {83}},\ \bibinfo
  {pages} {195131} (\bibinfo {year} {2011})}\BibitemShut {NoStop}%
\bibitem [{\citenamefont {Peng}\ \emph {et~al.}(2016)\citenamefont {Peng},
  \citenamefont {Yang}, \citenamefont {Perdew},\ and\ \citenamefont
  {Sun}}]{Peng2016}%
  \BibitemOpen
  \bibfield  {author} {\bibinfo {author} {\bibfnamefont {H.}~\bibnamefont
  {Peng}}, \bibinfo {author} {\bibfnamefont {Z.-H.}\ \bibnamefont {Yang}},
  \bibinfo {author} {\bibfnamefont {J.~P.}\ \bibnamefont {Perdew}}, \ and\
  \bibinfo {author} {\bibfnamefont {J.}~\bibnamefont {Sun}},\ }\bibfield
  {title} {\enquote {\bibinfo {title} {Versatile van der {W}aals density
  functional based on a meta-generalized gradient approximation},}\ }\href
  {\doibase 10.1103/PhysRevX.6.041005} {\bibfield  {journal} {\bibinfo
  {journal} {Phys. Rev. X}\ }\textbf {\bibinfo {volume} {6}},\ \bibinfo {pages}
  {041005} (\bibinfo {year} {2016})}\BibitemShut {NoStop}%
\bibitem [{\citenamefont {Monkhorst}\ and\ \citenamefont
  {Pack}(1976)}]{SPECIAL1976}%
  \BibitemOpen
  \bibfield  {author} {\bibinfo {author} {\bibfnamefont {H.~J.}\ \bibnamefont
  {Monkhorst}}\ and\ \bibinfo {author} {\bibfnamefont {J.~D.}\ \bibnamefont
  {Pack}},\ }\bibfield  {title} {\enquote {\bibinfo {title} {Special points for
  brillouin-zone integrations},}\ }\href {\doibase 10.1103/PhysRevB.13.5188}
  {\bibfield  {journal} {\bibinfo  {journal} {Phys. Rev. B}\ }\textbf {\bibinfo
  {volume} {13}},\ \bibinfo {pages} {5188} (\bibinfo {year}
  {1976})}\BibitemShut {NoStop}%
\end{thebibliography}

%

\end{document}